\documentclass[12 pt]{article}
\usepackage[left=1in, right=1in, top=0.75in, bottom=0.75in]{geometry}
\usepackage[utf8]{inputenc}
\usepackage{textcomp}
\usepackage{setspace}
\usepackage{graphicx}
\usepackage{chemformula}
\usepackage{authblk}
\usepackage{booktabs}
\usepackage{gensymb}
\usepackage{amsmath}
\usepackage{parskip}
\usepackage{caption}
\usepackage{titlesec}
\usepackage{wrapfig}

\titlespacing{\section}{0pt}{12pt plus 2pt minus 2pt}{12pt plus 2pt minus 2pt}

\usepackage{sectsty}
\sectionfont{\fontsize{14}{16}\selectfont}
\subsectionfont{\fontsize{12}{13}\selectfont}

\usepackage[backend=biber, style=chem-acs, articletitle=true, sorting=none, maxbibnames=99]{biblatex}
\newcommand{\CrxNbS}{\ch{Cr_{1/3}NbS2}}
\newcommand{\CrxTaS}{\ch{Cr_{1/3}TaS2}}
\newcommand{\CrxMS}{\ch{Cr_{1/3}$M$S2}}

\title{\Large\textbf{Comparative Electronic Structures of the Chiral Helimagnets \CrxNbS\ and \CrxTaS}}

\author[1,$\dagger$]{Lilia S.\ Xie}
\author[1,$\dagger$]{Oscar Gonzalez}
\author[2]{Kejun Li}
\author[3,4]{Matteo Michiardi}
\author[5]{Sergey Gorovikov}
\author[6]{Sae Hee Ryu}
\author[1]{Shannon S.\ Fender}
\author[5]{Marta Zonno}
\author[6,7]{Na Hyun Jo}
\author[3,4]{Sergey Zhdanovich}
\author[6]{Chris Jozwiak}
\author[6]{Aaron Bostwick}
\author[1]{Samra Husremovi\'{c}}
\author[1]{Matthew P.\ Erodici}
\author[1]{Cameron Mollazadeh}
\author[3,4]{Andrea Damascelli}
\author[6]{Eli Rotenberg}
\author[8,9]{Yuan Ping}
\author[1,10,*]{D.\ Kwabena Bediako}
\affil[1]{Department of Chemistry, University of California, Berkeley, CA 94720, USA}
\affil[2]{Department of Physics, University of California, Santa Cruz, CA, 95064, USA}
\affil[3]{Quantum Matter Institute, University of British Columbia, Vancouver, BC V6T 1Z4, Canada}
\affil[4]{Department of Physics and Astronomy, University of British Columbia, Vancouver, BC V6T 1Z1, Canada}
\affil[5]{Canadian Light Source, Inc., 44 Innovation Boulevard, Saskatoon SK S7N 2V3, Canada}
\affil[6]{Advanced Light Source, Lawrence Berkeley National Laboratory, Berkeley, CA 94720, USA}
\affil[7]{Department of Physics, University of Michigan, Ann Arbor, Michigan 48109, USA}
\affil[8]{Department of Chemistry and Biochemistry, University of California, Santa Cruz, CA, 95064, USA}
\affil[9]{Department of Materials Science and Engineering, University of Wisconsin, Madison, WI, 53706, USA}
\affil[10]{Chemical Sciences Division, Lawrence Berkeley National Laboratory, Berkeley, CA 94720, USA}
\affil[*]{Correspondence to: bediako@berkeley.edu}
\affil[$\dagger$]{These authors contributed equally to this work}

\date{}
\begin{document}
\maketitle

\doublespacing

\newrefsection

\section*{Abstract}

Magnetic materials with noncollinear spin textures are promising for spintronic applications. To realize practical devices, control over the length and energy scales of such spin textures is imperative. The chiral helimagnets \CrxNbS\ and \CrxTaS\ exhibit analogous magnetic phase diagrams with different real-space periodicities and field dependence, positioning them as model systems for studying the relative strengths of the microscopic mechanisms giving rise to exotic spin textures. Here, we carry out a comparative study of the electronic structures of \CrxNbS\ and \CrxTaS\ using angle-resolved photoemission spectroscopy and density functional theory. We show that bands in \CrxTaS\ are more dispersive than their counterparts in \CrxNbS\ and connect this result to bonding and orbital overlap in these materials. We also unambiguously distinguish exchange splitting from surface termination effects by studying the dependence of their photoemission spectra on polarization, temperature, and beam size. We find strong evidence that hybridization between intercalant and host lattice electronic states mediates the magnetic exchange interactions in these materials, suggesting that band engineering is a route toward tuning their spin textures. Overall, these results underscore how the modular nature of intercalated transition metal dichalcogenides translates variation in composition and electronic structure to complex magnetism.

\section*{Introduction}

Next-generation spintronic devices utilize the spin degree of freedom to store information. \supercite{fert2013,parkin2015} Magnetic materials in which spins order in topologically protected quasiparticles, such as skyrmions or magnetic solitons, are promising platforms for realizing such devices.\supercite{togawa2016,tokura2021} These chiral spin textures can be manipulated with currents and magnetic fields, which is appealing for various applications in memory, logic, and unconventional computing.\supercite{tey2022} For practical spintronic devices, optimizing the energy and length scales of the spin textures is important: stability at operationally accessible temperatures and fields as well as high density in thin-film architectures are broadly desirable. Strategies to control the microscopic mechanisms that give rise to complex magnetism are thus needed. In terms of materials design, this can be broadly achieved by tailoring the interactions between spin centers as directed by their spatial arrangements and coordination environments.

The chiral helimagnets \CrxNbS\ and \CrxTaS\ are especially well-suited for device schemes implementing noncollinear spin textures because of their anisotropic layered structures, which are compatible with thin-film architectures.\supercite{togawa2015,yamasaki2017,wang2017,zhang2022,osorio2022} In these materials, the $S=3/2$ Cr$^{3+}$ centers occupy pseudo-octahedral sites between layers of $2H$-\ch{NbS2} or $2H$-\ch{TaS2},\supercite{parkin1980a,parkin1980} forming a $\sqrt{3} \times \sqrt{3}$ superlattice.\supercite{rouxel1971} They exhibit easy-plane ferromagnetic (FM) behavior with chiral magnetic ordering out-of-plane: the Cr superlattice breaks the inversion symmetry of the transition metal dichalcogenide (TMD) host lattice along the crystallographic $c$-axis, giving rise to a Dzyaloshinskii--Moriya (DM) interaction, also known as antisymmetric exchange.\supercite{dzyaloshinsky1958,moriya1960} The DM interaction favors out-of-plane spin canting, which competes with FM exchange to produce one-dimensional helical spin textures that propagate along [001]. Importantly, the application of an in-plane magnetic field creates a chiral soliton lattice (CSL) phase with tunable periodicities up to a critical field, $H_\mathrm{c}$, above which a forced ferromagnetic (FFM) state is observed.\supercite{miyadai_magnetic_1983,togawa_chiral_2012-1,ghimire2013,zhang2021,obeysekera2021,du2021} Both \CrxNbS\ and \CrxTaS\ have Curie temperatures, $T_{\mathrm{C}}$, well above 100~K, and nanoscale soliton wavelengths tunable with fields of 1.5~T or less, thus providing a richly accessible phase space for manipulating chiral spin textures.

Although the magnetic phase diagrams for \CrxNbS\ and \CrxTaS\ are qualitatively analogous, the periodicities and stabilities of their magnetic solitons differ somewhat. Literature reports have established that \CrxTaS\ consistently exhibits a higher $T_{\mathrm{C}}$,\supercite{togawa_chiral_2012-1,ghimire2013,du2021,kousaka2022,meng2023}, higher $H_\mathrm{c}$,\supercite{ghimire2013,zhang2021,obeysekera2021,meng2023} and a shorter soliton wavelength than the Nb analogue.\supercite{togawa_chiral_2012-1,kousaka2016,zhang2021,du2021} These observations imply that changing the host lattice from \ch{NbS2} to \ch{TaS2} alters the relative strengths of magnetic coupling among Cr centers, manifesting in quantitative changes to their magnetic phase diagrams. However, the origin of magnetic exchange interactions in these materials is still a matter of debate,\supercite{sirica2016,sirica2021,qin2022,hicken2022} and comparative studies have been scant.\supercite{hicken2022} A detailed investigation of the electronic structures of both \CrxNbS\ and \CrxTaS\ is thus motivated by the fact that these materials are natural platforms for studying how the length and energy scales of chiral spin textures can be tuned through materials chemistry. 

Herein, we present a comprehensive investigation of the electronic structures of \CrxNbS\ and \CrxTaS\ using angle-resolved photoemission spectroscopy (ARPES) and density functional theory (DFT) calculations. We show that the Ta analogue has more dispersive bands, consistent with greater orbital overlap in the case of Ta, and discuss implications for their magnetic properties and the rational design of chiral helimagnets. Using polarization-dependent ARPES and orbital-projected DFT calculations, we assign the parity and orbital character of bands, finding excellent agreement between theory and experiment with the exception of additional band splitting near the Fermi level observed in ARPES but not predicted in DFT. ARPES data collected with smaller beam sizes reveal spatial variation in this band splitting, consistent with different surface terminations on the as-cleaved samples. Thus, we distinguish exchange splitting from surface vs.\ bulk splitting for the first time in these materials. Our findings establish a high degree of similarity in the electronic structures of \CrxNbS\ and \CrxTaS\ and highlight the relevance of their polar layered nature in interpreting surface-sensitive spectroscopic investigations. More generally, these results suggest that careful band structure engineering and Fermi level tuning may prove to be fruitful avenues toward optimizing the spin textures in intercalated TMDs.

\begin{figure*}[htbp]
\centering
\includegraphics[width=6in]{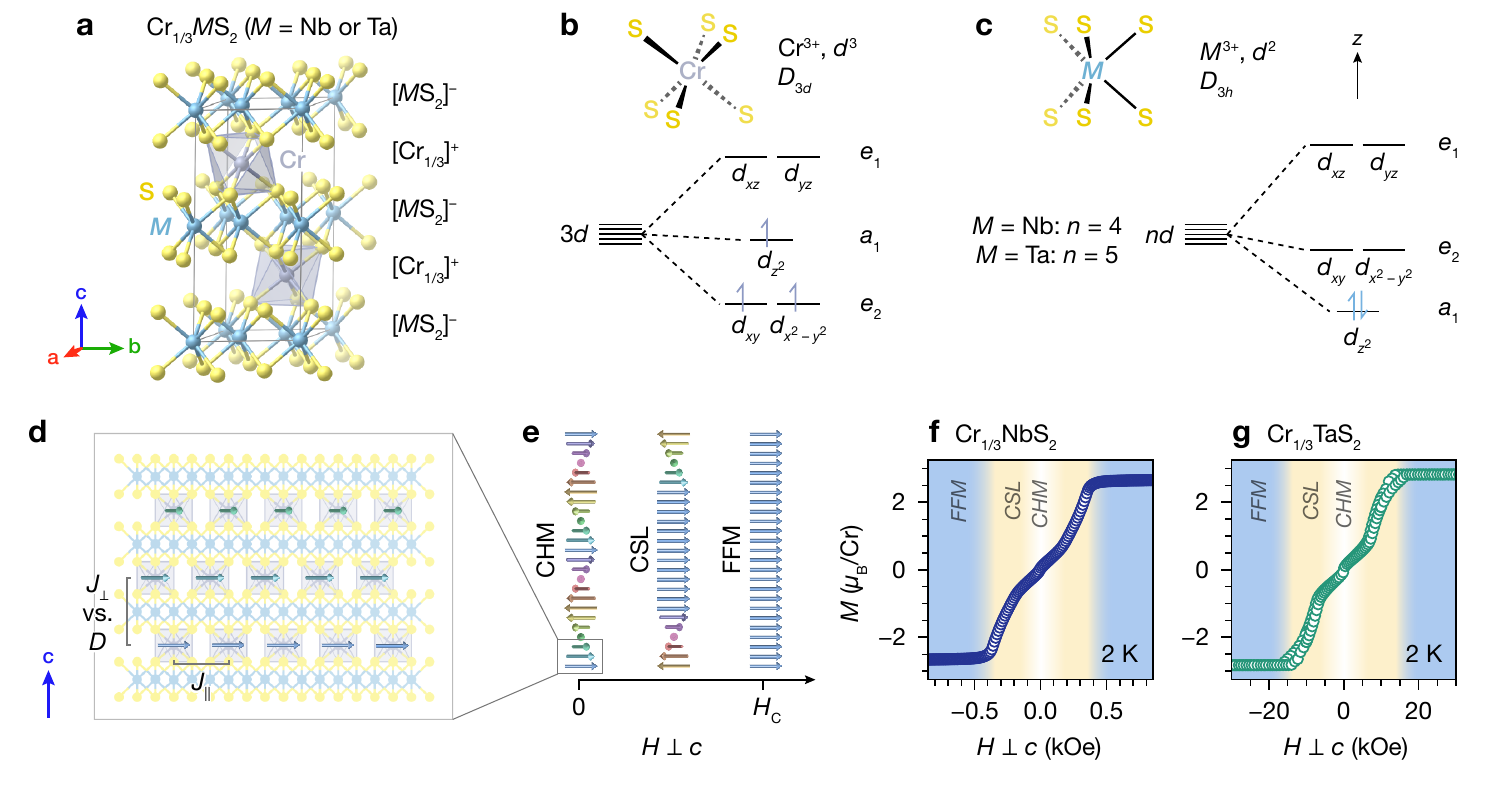}
\caption{\label{fig:overview}(a) Crystal structure of \CrxMS, showing formal charges for the \ch{$M$S2} and Cr layers from a simple electron-counting picture. (b) and (c) Qualitative $d$-orbital splitting diagrams for isolated Cr and $M$ centers from the local ligand field in \CrxMS. (d) Schematic illustration of the magnetic structure of \CrxMS\ in the chiral helimagnetic (CHM) state. (e) Schematic representations of spin textures evolving CHM to chiral soliton lattice (CSL) to forced ferromagnetic (FFM) states with increasing applied magnetic field $H \perp c$. (f) and (g) $M(H)$ data for \CrxNbS\ and \CrxTaS, respectively, showing transitions between CHM, CSL, and FFM states.}
\end{figure*}

First, we briefly outline the electronic and magnetic properties of \CrxMS\ ($M$ = Nb or Ta) as established in the existing literature. According to a simple electron counting scheme, these compounds can be considered as alternating layers of \ch{[Cr_{1/3}]+} and \ch{[$M$S2]-}. The intercalant layers consist of \ch{Cr^{3+}} centers occupying 1/3 of the trigonally distorted pseudo-octahedral interstitial sites between layers of $2H$-\ch{$M$S2} ($M$ = Nb or Ta). These intercalant layers donate one electron per formula unit to the \ch{$M$S2} host lattice layers (Figure \ref{fig:overview}a). The qualitative local $d$-orbital splitting diagrams for the \ch{Cr^{3+}} ($D_{3d}$) and \ch{$M$^{3+}} ($D_{3h}$) centers are shown in Figure \ref{fig:overview}b--c.\supercite{xie_structure_2022} The electronic structure of the periodic solids are more complex, and in reality, the half-filled TMD bands have both $d_{z^2}$ and $d_{xy}$/$d_{x^2-y^2}$ character.\supercite{mattheiss1973,yee1991,whangbo1992} Nevertheless, this simplified picture captures (1) charge transfer from the intercalant species to the highest-lying $M$ $d$ bands of the host lattice, and (2) the inherently polar nature of these layered intercalation compounds.

The qualitative magnetic properties of \CrxNbS\ and \CrxTaS\ below $T_{\mathrm{C}}$ are summarized in Figure \ref{fig:overview}d--e. Within each \ch{[Cr_{1/3}]+} layer, the Cr spins exhibit FM coupling through an in-plane exchange constant, $J_{\parallel}$. Between adjacent \ch{[Cr_{1/3}]+} layers, FM coupling through an out-of-plane exchange constant, $J_{\perp}$, competes energetically with spin canting through a DM interaction term, $D$. At zero field, this results in a continuous helical arrangement of spins, or a CHM ground state, with the magnetic soliton wavelength determined by the ratio of $J_{\perp}$ and $D$.\supercite{chapman2014spin,aczel2018} With increasing $H \perp c$, FM regions aligned with the field grow, effectively unwinding the CHM state to create the CSL phase, in which the distance separating adjacent solitons is a function of the magnitude of $H$. Finally, with fields larger than $H_\mathrm{c}$, an FFM state with saturated magnetization is obtained.\supercite{miyadai_magnetic_1983,togawa_chiral_2012-1,ghimire2013,han2017,zhang2021,obeysekera2021,meng2023}

In this study, we investigate the electronic structure of \CrxNbS\ and \CrxTaS\ in a comparative context to tease out differences between the two compounds and connect these to their magnetic phase diagrams. To do so, we grew and characterized single crystals, verified their chiral spin textures with magnetometry, carried out a comprehensive suite of ARPES measurements, and conducted DFT band structure calculations, as detailed below.

\section*{Results}

\subsection*{Synthesis, Structure, and Magnetism}

Single crystals of \CrxNbS\ and \CrxTaS\ were grown via chemical vapor transport using iodine as a transport agent. X-ray diffraction confirmed that both materials crystallize in the noncentrosymmetric space group $P6_322$, with the Cr centers forming a $\sqrt{3} \times \sqrt{3}$ superlattice (Figure \ref{fig:structure} and Tables \ref{tab:crystal}--\ref{tab:CrxTaS2}). \CrxNbS\ exhibits a slightly larger in-plane lattice parameter and smaller out-of-plane lattice parameter ($a$ = 5.7400(7) \AA\ and $c$ = 12.1082(14) \AA) compared to \CrxTaS\ ($a$ = 5.7155(5) \AA\ and $c$ = 12.1751(12) \AA). Raman spectroscopy revealed sharp vibrational modes associated with the $\sqrt{3} \times \sqrt{3}$ superlattices\supercite{fan2021} (Figure \ref{fig:raman}), and energy dispersive X-ray spectroscopy indicated Cr:Nb and Cr:Ta ratios of 0.33(1):1 (Figures \ref{fig:EDS_Nb} and \ref{fig:EDS_Ta}). 

The metamagnetic transitions across these states with applied magnetic field are observed in the $M(H)$ data for single crystals of \CrxNbS\ and \CrxTaS\ shown in Figure \ref{fig:overview}f--g, confirming the characteristic spin textures in our samples.\supercite{miyadai_magnetic_1983,zhang2021,obeysekera2021} Both compounds exhibit similar saturation moments (2.7 $\mu_{\mathrm{B}}$/Cr for \CrxNbS\ and 2.8 $\mu_{\mathrm{B}}$/Cr for \CrxTaS), close to the expected spin-only value of 3 $\mu_{\mathrm{B}}$/Cr. The analogous transitions are observed at fields more than an order of magnitude larger for \CrxTaS\ than \CrxNbS, with $H_\mathrm{c}$ values of about 0.45 mT for \CrxNbS\ and 16 mT for \CrxTaS. This is consistent with shorter soliton wavelengths in the Ta analogue.\supercite{zhang2021,obeysekera2021,du2021} The $M(T)$ data show pronounced peaks at 110 and 133 K for \CrxNbS\ and \CrxTaS, respectively, corresponding to the onset of chiral helimagnetism below these temperatures (Figures \ref{fig:MT_Nb}--\ref{fig:MT_Ta}).

\subsection*{Superlattice Effects on Electronic Structure}

\begin{figure*}[htbp]
\centering
\includegraphics[width=3.33in]{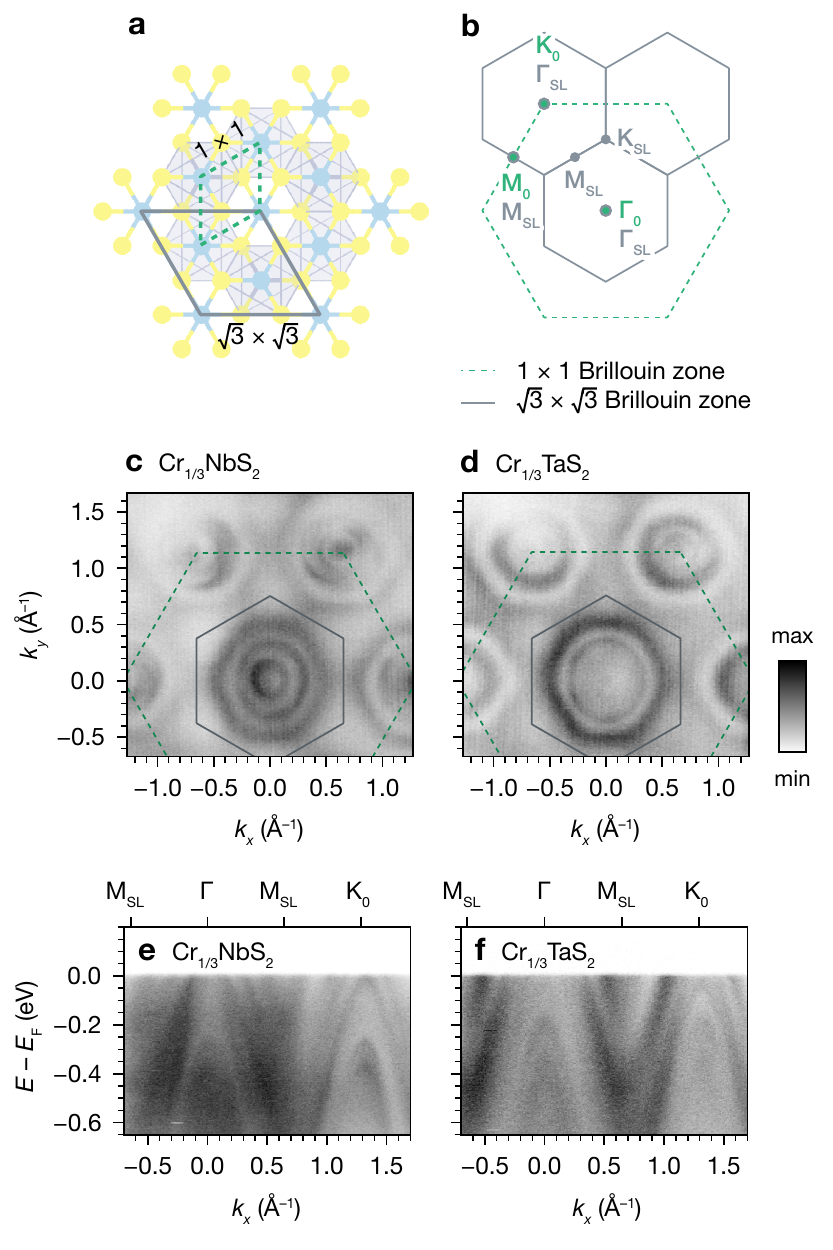}
\caption{\label{fig:folding}(a) Real-space crystal structure of \CrxMS\ ($M$ = Nb or Ta) viewed along the crystallographic $c$-axis, with overlaid unit cells for the $1 \times 1$ primitive \ch{$M$S2} lattice (dashed green) and the $\sqrt{3} \times \sqrt{3}$ Cr superlattice (solid gray). (b) Surface Brillouin zones for the $1 \times 1$ primitive lattice and $\sqrt{3} \times \sqrt{3}$ superlattice. (c) and (d) ARPES Fermi surfaces of \CrxNbS\ and \CrxTaS\ with the dashed and solid overlaid lines corresponding to the primitive and  $\sqrt{3} \times \sqrt{3}$ superlattice Brillouin zones, respectively. (e) and (f) ARPES band dispersions for \CrxNbS\ and \CrxTaS\ along the $\Gamma$--$\mathrm{K}_0$ direction, showing folding of features from $\Gamma$ to $\mathrm{K}_0$ and vice versa (18 K, $h\nu$ = 79 eV).}
\end{figure*}

After obtaining structural and magnetic evidence of highly ordered $\sqrt{3} \times \sqrt{3}$ Cr superlattices in \CrxNbS\ and \CrxTaS, we sought to investigate their influence on the electronic structure of these materials. Figure \ref{fig:folding}a illustrates the real-space $1 \times 1$ primitive unit cell for the host lattice TMD and the $\sqrt{3} \times \sqrt{3}$ superlattice unit cell for the intercalated compounds along [001]. The $\sqrt{3} \times \sqrt{3}$ unit cell is rotated by 30\textdegree\ compared to the $1 \times 1$ unit cell. In reciprocal space, the $\sqrt{3} \times \sqrt{3}$ superlattice defines a smaller Brillouin zone that is likewise rotated by 30\textdegree\ relative to the primitive Brillouin zone (Figure \ref{fig:folding}b). To probe the electronic effects of Cr intercalation, we first examined the symmetries of the experimental Fermi surfaces and band dispersions of \CrxNbS\ and \CrxTaS\ using ARPES.

As shown in Figure \ref{fig:folding}c--d, the Fermi surfaces of both \CrxNbS\ and \CrxTaS\ below $T_{\mathrm{C}}$ ($h\nu$ = 79 eV) display multiple nested barrels around $\Gamma$ and $\mathrm{K}$ of the primitive Brillouin zone, which is indicated by the dashed green hexagons. Notably, six-fold symmetry is clearly observed around the primitive $\mathrm{K}$ (denoted as $\mathrm{K_0}$), in contrast with three-fold symmetry around $\mathrm{K}$ of the host lattice materials $2H$-\ch{NbS2} and $2H$-\ch{TaS2}.\supercite{elyoubi2021,zhao2017} Additionally, in the intercalated materials, three-fold symmetry is introduced at $\mathrm{K}$ of the $\sqrt{3} \times \sqrt{3}$ superlattice Brillouin zone (denoted as $\mathrm{K_{SL}}$), which is indicated by the solid gray hexagons in Figure \ref{fig:folding}c--d. Hence, the Fermi surfaces of \CrxNbS\ and \CrxTaS\ display the expected symmetries associated with reconstruction and band folding from the $\sqrt{3} \times \sqrt{3}$ Cr superlattice.

The ARPES dispersions show clear evidence of band folding as well (Figure \ref{fig:folding}e--f). Cuts along the $\Gamma$--$\mathrm{K_0}$ direction show the same features at both $\Gamma$ and $\mathrm{K_0}$: both materials display several nested hole pockets and parabolic bands below $E_{\mathrm{F}}$. In contrast, for the host TMDs $2H$-\ch{NbS2} and $2H$-\ch{TaS2}, the bands crossing $E_{\mathrm{F}}$ have different dispersions and energies at $\Gamma$ and $\mathrm{K}$. In the Cr-intercalated materials, the $\sqrt{3} \times \sqrt{3}$ superlattice folds the primitive lattice $\Gamma$ to $\mathrm{K}$ and vice versa, as they both become $\Gamma$ of the superlattice Brillouin zone (denoted as $\mathrm{\Gamma_{SL}}$ in Figure \ref{fig:folding}b). Thus, the presence of the same features at both $\Gamma$ and $\mathrm{K_0}$ in the ARPES of the Cr-intercalated materials is consistent with $\sqrt{3} \times \sqrt{3}$ superlattice band folding.

The band folding in the ARPES data reveals that the $\sqrt{3} \times \sqrt{3}$ Cr superlattice potential is strong in both \CrxNbS\ and \CrxTaS. Broadly, this electronic reconstruction is in line with previous literature reports on \CrxNbS,\supercite{sirica2016,qin2022} as well as other intercalated TMDs with $\sqrt{3} \times \sqrt{3}$ transition metal superlattices.\supercite{tanaka2022,yang2022,popcevic2022,edwards2023} The features observed in both materials are qualitatively similar; however, at a glance, the hole pockets in \CrxTaS\ appear to be larger than those found in \CrxNbS. To contextualize differences in the experimental electronic structures of \CrxNbS\ and \CrxTaS, we turned to DFT calculations and quantitative analysis of their band dispersions.

\subsection*{Relative Band Dispersions}

\begin{figure*}[htbp]
\centering
\includegraphics[width=3.33in]{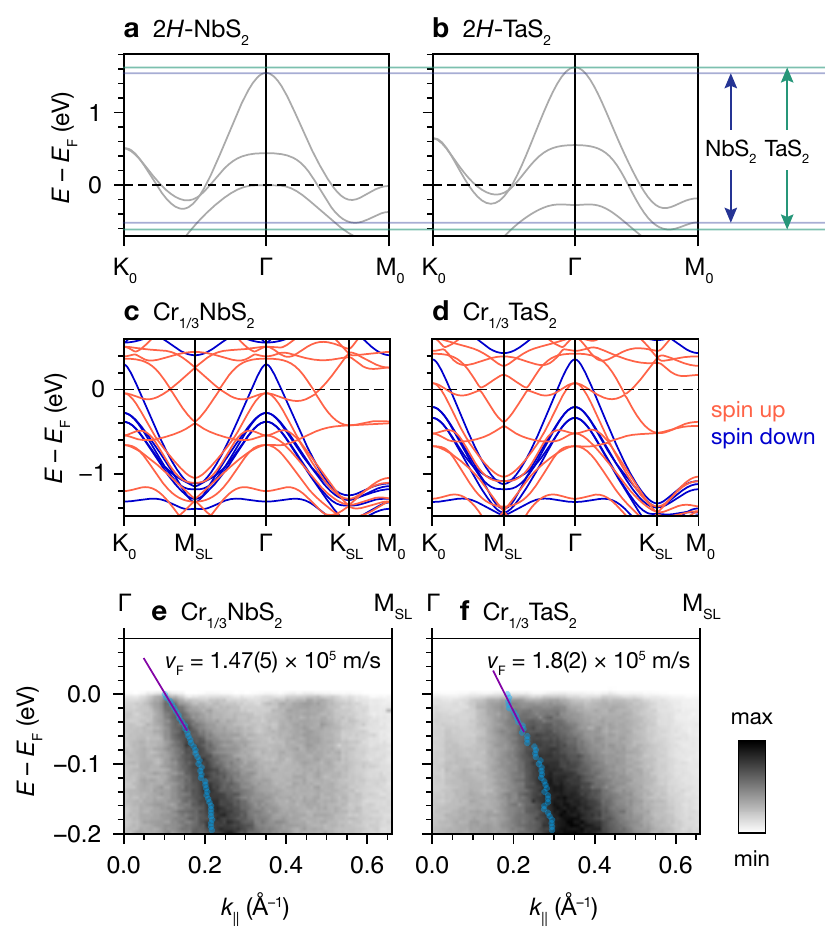}
\caption{\label{fig:dispersion}(a) and (b) DFT band structures of $2H$-\ch{NbS2} and $2H$-\ch{TaS2}, with maxima and minima of the bands crossing $E_{\mathrm{F}}$ indicated by solid navy and mint lines, respectively. (c) and (d) Spin-polarized band structures for \CrxNbS\ and \CrxTaS\ in the FM state, with spin up and spin down bands indicated in red and blue, respectively. (e) and (f) ARPES dispersions of \CrxNbS\ and \CrxTaS\ (18 K, $h\nu$ = 46 eV), with blue circles indicating the peak center positions of the most intense feature from MDC analysis. The Fermi velocities, $v_{\mathrm{F}}$, are obtained from linear fits to the centers between 0 and $-50$ meV.}
\end{figure*}

To understand the relative differences between the band structures of \CrxNbS\ and \CrxTaS, we started by comparing the host lattice materials, $2H$-\ch{NbS2} and $2H$-\ch{TaS2}. DFT band structure calculations of $2H$-\ch{NbS2} and $2H$-\ch{TaS2} show that the bands crossing $E_{\mathrm{F}}$ in $2H$-\ch{TaS2} are more dispersive compared to the analogous bands in $2H$-\ch{NbS2}. This can be clearly visualized by comparing the relative spread of the maxima and minima of these respective bands, as illustrated in Figure \ref{fig:dispersion}a--b: the more dispersive bands in $2H$-\ch{TaS2} have a higher-energy maximum and lower-energy minimum compared to $2H$-\ch{NbS2}. These bands have predominantly Nb or Ta $d_{z^2}$ and $d_{xy}$/$d_{x^2-y^2}$ character, with additional contribution from S $p$ states.\supercite{zhao2017,elyoubi2021}

Next, we calculated the band structures of the Cr-intercalated materials and compared the results to our ARPES data. Due to the surface-sensitive nature of ARPES, we do not expect to experimentally resolve signatures of the CHM state, i.e.\ out-of-plane spin textures with length scales on the order of tens of nm. Hence, we use spin-polarized band structure calculations of \CrxNbS\ and \CrxTaS\ in their FM states, with the magnetization vector along [100], as proxies for the electronic structure near the surface (Figure \ref{fig:dispersion}c--d and Figures \ref{fig:bandstruct_crnbs2_vs_U}--\ref{fig:slab}). Three distinct changes are evident in the DFT band structures of \CrxNbS\ and \CrxTaS\ compared to the host lattices: (1) folding due to the $\sqrt{3} \times \sqrt{3}$ superlattice potential, (2) raising of $E_{\mathrm{F}}$ due to electron transfer from Cr to the host lattice, and (3) introduction of new bands crossing $E_{\mathrm{F}}$ due to Cr--Nb or Cr--Ta hybridization and FM exchange splitting.

Although the Cr-intercalated materials have more complex electronic structures than the host lattices, DFT calculations show that the Ta analogue again has more dispersive bands than the Nb analogue. The amount of charge transfer from Cr to the host lattice appears to be very similar for both materials, as shown by the calculated and experimental magnetic moments (Tables \ref{tab:magn_crnbs} and \ref{tab:magn_crtas}). Thus, the shift of $E_{\mathrm{F}}$ upon intercalation is almost identical. This results in larger hole pockets around $\Gamma$ and $\mathrm{K_0}$ in \CrxTaS\ than \CrxNbS\, and an extra spin-up band crossing $E_{\mathrm{F}}$ at $\Gamma$ and $\mathrm{K}_0$ in \CrxTaS. Notably, the ARPES dispersions of \CrxNbS\ and \CrxTaS\ at 18 K ($h\nu$ = 46 eV) show clearly that the most intense hole pocket around $\Gamma$ in the $\Gamma$--$\mathrm{M_{SL}}$ direction is considerably larger at $E = E_{\mathrm{F}}$ in \CrxTaS\ compared to \CrxNbS\ (Figure \ref{fig:dispersion}e--f), with Fermi wavevectors, $k_{\mathrm{F}}$, of 0.10 \AA\ for \CrxNbS\ and 0.19 \AA\ for \CrxTaS. By fitting the momentum distribution curves (MDCs) to Lorentzians between 0 and $-50$ meV, we extracted Fermi velocities, $v_{\mathrm{F}}$, of $1.47(5) \times 10^5$ m/s for \CrxNbS\ and $1.8(2) \times 10^5$ m/s for \CrxTaS---thus experimentally quantifying the relative band dispersions between the two systems. The larger experimental $v_{\mathrm{F}}$ for the Ta analogue mirrors the relative trends from the DFT band structures.

\subsection*{Orbital Character Assignments}

\begin{figure*}[htbp]
\centering
\includegraphics[width=3.33in]{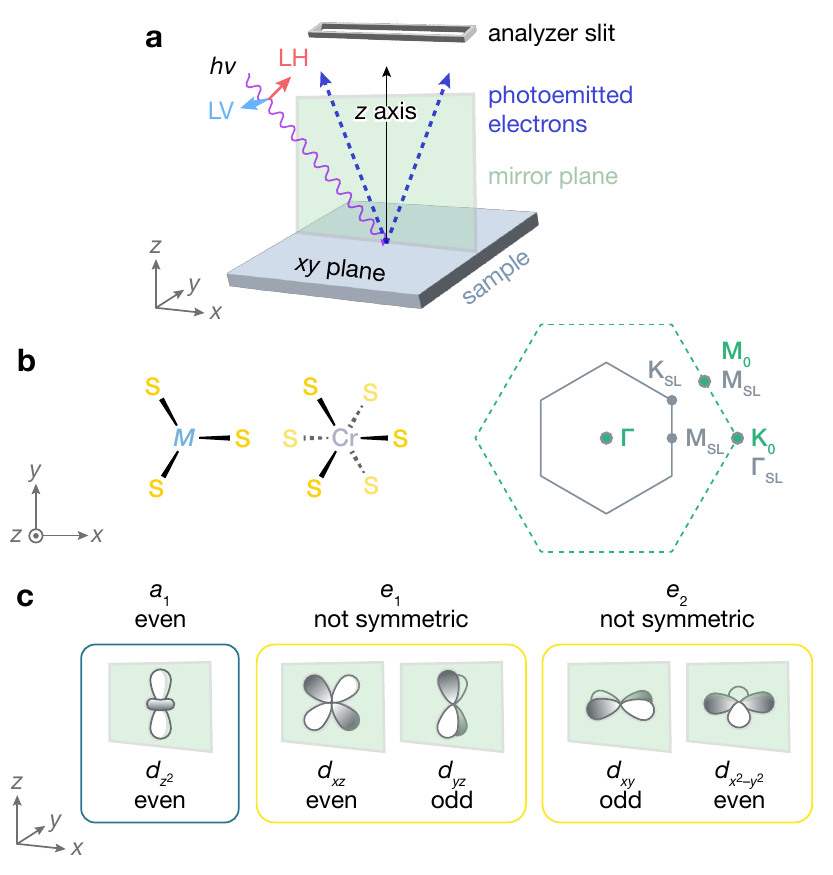}
\caption{\label{fig:parity} (a) Geometry of ARPES data collection for a horizontal analyzer slit aligned with the $xz$ scattering plane of the sample. (b) Surface Brillouin zones for the primitive lattice (dotted green) and $\sqrt{3} \times \sqrt{3}$ superlattice (solid gray), along with real space projections of local coordination environments for $M$ = Nb or Ta and Cr. (c) Symmetries of $d$ orbitals for $M$ = Nb or Ta and Cr for the scattering plane defined as the $xz$ plane aligned with the $\Gamma$--$\mathrm{K}_0$ direction.}
\end{figure*}

To gain insight into the orbital character of the bands, we studied their polarization dependence in ARPES. For the photoemission process, the matrix element term can be described by
\[ | M^{\mathbf{k}}_{f,i}|^2 \propto | \langle \phi^{\mathbf{k}}_f | \hat{\epsilon} \cdot \mathbf{r} | \phi^{\mathbf{k}}_i \rangle |^2 \]
where $\hat{\epsilon}$ is the unit vector along the polarization direction of the light.\supercite{damascelli2003} The final state wavefunction of the photoelectron, $\phi^{\mathbf{k}}_f$, can be described by a plane wave state, $e^{i\mathbf{kr}}$, with even parity with respect to the mirror plane defined by the analyzer slit and the normal to the sample surface (Figure \ref{fig:parity}a). To obtain a nonvanishing matrix element, $\hat{\epsilon}$ must be even (odd) for an even (odd) initial state wavefunction, $\phi^{\mathbf{k}}_i$. Based on the symmetry operations of space group $P6_322$ (point group $D_6$), and taking $z$ to be parallel to the crystallographic $c$-axis, we expect the even $a_1$ ($d_{z^2}$) states of both Cr and Nb or Ta to be visible with linear horizontal (LH) polarized light (even $\hat{\epsilon}$) but not linear vertical (LV) polarized light (odd $\hat{\epsilon}$).

The $e_1$ ($d_{xz}$, $d_{yz}$) and $e_2$ ($d_{xy}$, $d_{x^2-y^2}$) sets are not symmetric overall with respect to any scattering plane containing the sample surface normal. We illustrate this by considering a horizontal analyzer slit aligned to the $\Gamma$--$\mathrm{K}_0$ direction, and defining $x$ as parallel to the crystallographic $a$-axis of the superlattice unit cell. The resulting scattering plane is the $xz$ plane (Figure \ref{fig:parity}a) and contains $M$--S and Cr--S bonds (Figure \ref{fig:parity}b). As shown in Figure \ref{fig:parity}c, the $d_{xz}$ and $d_{x^2-y^2}$ orbitals are even with respect to the $xz$ plane, but the $d_{yz}$ and $d_{xy}$ orbitals (the other components of the $e_1$ and $e_2$ sets) are odd. Thus, the $e_1$ and $e_2$ sets are not symmetric collectively, and may be visible with both LH and LV polarization.

\begin{figure*}[htbp]
\centering
\includegraphics[width=\textwidth]{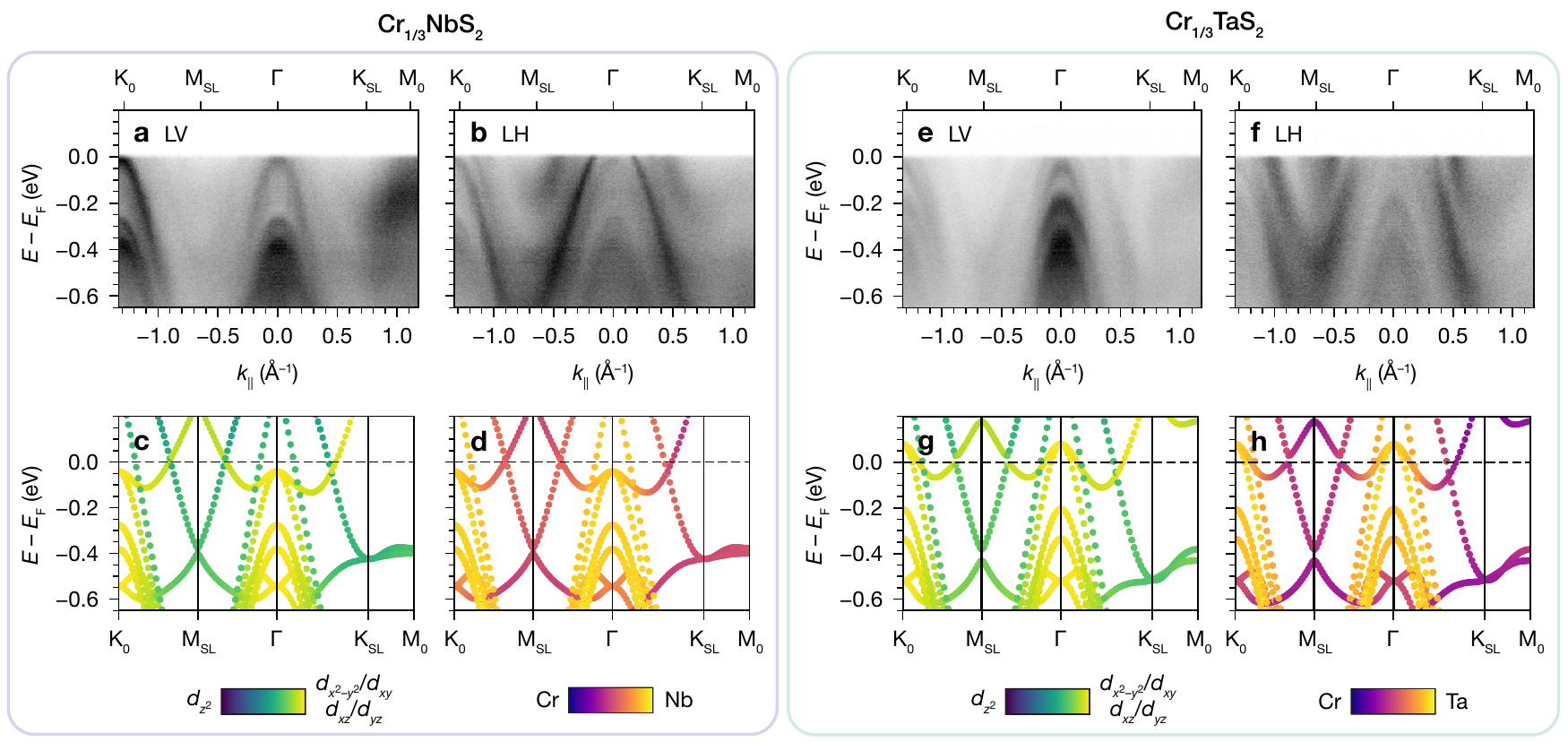}
\caption{\label{fig:polarization} (a) and (b) ARPES band dispersions for \CrxNbS\ with linear vertical (LV) and linear horizontal (LH) polarized photons, respectively (18 K, $h\nu$ = 79 eV). (c) and (d) DFT orbital-projected band structures of \CrxNbS\ in the FM state, showing in-plane vs.\ out-of-plane character, and Cr vs.\ Nb character, respectively. (e--h) The same as (a--d) for \CrxTaS, and Cr vs.\ Ta character in (h).}
\end{figure*}

ARPES data of \CrxNbS\ measured with LV polarization (Figure \ref{fig:polarization}a) show stronger intensity from the innermost parabolic bands centered at $\Gamma$ and especially $\mathrm{K_0}$. In contrast, with LH polarization (Figure \ref{fig:polarization}b), the sharp outermost dispersive bands around $\Gamma$ and $\mathrm{K_0}$ are more prominent, as well as two sets of more diffuse electron pockets with minima at $\mathrm{M_{SL}}$ and $\mathrm{K_{SL}}$. To compare with the polarization-dependent ARPES data, we plotted the orbital-projected DFT band structure as a function of even ($d_{z^2}$) vs.\ not symmetric ($d_{xy}$/$d_{x^2-y^2}$ and $d_{xz}$/$d_{yz}$) states in Figure \ref{fig:polarization}c. At $\Gamma$/$\mathrm{K_0}$, the innermost parabolic bands have predominantly $d_{xy}$/$d_{x^2-y^2}$ and $d_{xz}$/$d_{yz}$ character, whereas the outermost dispersive bands and electron pockets have more $d_{z^2}$ character. These parities are qualitatively consistent with the experimentally observed polarization dependence. The DFT band structure as a function of Cr vs.\ Nb character (\ref{fig:polarization}d) indicates that all of the parabolic bands at $\Gamma$/$\mathrm{K_0}$ are predominantly Nb-derived, while the electron pockets with minima at $\mathrm{M_{SL}}$ and $\mathrm{K_{SL}}$ are composed of mixed Cr and Nb states. The polarization dependence of the host lattice bands more visible in LV polarization is consistent with $d_{xy}$/$d_{x^2-y^2}$ states folded to $\Gamma$ from $\mathrm{K_0}$ by the $\sqrt{3} \times \sqrt{3}$ superlattice potential.\supercite{elyoubi2021}

The polarization-dependent ARPES data for \CrxTaS\ are similar to those for \CrxNbS. The innermost bands at $\Gamma$ and $\mathrm{K_0}$ are more prominent in LV polarization (Figure \ref{fig:polarization}e), whereas the outer bands around $\Gamma$ and $\mathrm{K_0}$ and more diffuse electron pockets with minima at $\mathrm{M_{SL}}$ and $\mathrm{K_{SL}}$ are more intense in LH polarization (Figure \ref{fig:polarization}f). The orbital-projected DFT band structure reveals analogous parities to the Nb analogue (Figure \ref{fig:polarization}g) and similar atomic parentage (Figure \ref{fig:polarization}h), albeit with more Cr character in the vicinity of $E_\mathrm{F}$. Less Cr--Ta hybridization may be an effect of the slightly longer $c$ lattice parameter in \CrxTaS.

\begin{figure*}[htbp]
\centering
\includegraphics[width=5.3in]{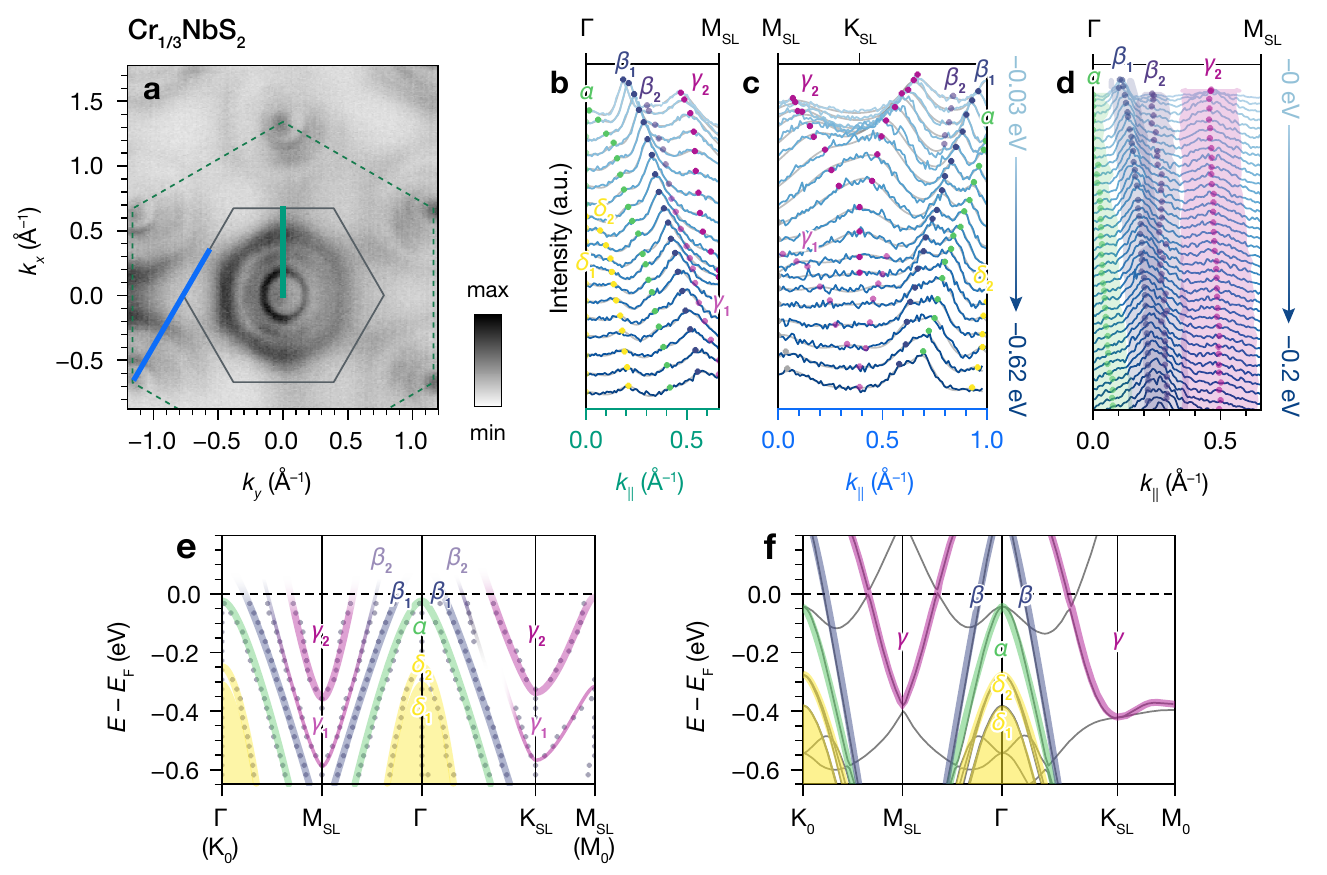}
\caption{\label{fig:MDCs} (a) ARPES Fermi surface of \CrxNbS. Dashed and solid overlaid lines indicate the primitive and $\sqrt{3} \times \sqrt{3}$ superlattice Brillouin zones, respectively. (b) and (c) MDCs along the $\Gamma$--$\mathrm{M_{SL}}$ and $\mathrm{M_{SL}}$--$\mathrm{K_{SL}}$ directions (cuts indicated by bold blue and teal lines in (a)). Gray lines are multi-Lorentzian fits, and colored circles indicate peak center positions, with band assignments labeled. (d) MDCs along the $\Gamma$--$\mathrm{M_{SL}}$ direction (18 K, $h\nu$ = 46 eV). Shaded regions correspond to the full width at half maximum values determined by multi-Lorentzian fits. (e) Sketch of the proposed band structure of \CrxNbS\ derived from the MDC fits (peak center positions shown by gray circles). (f) DFT band structure of \CrxNbS, with corresponding band assignments from MDC analysis indicated by colored overlays.}
\end{figure*}

For a more quantitative enumeration of the bands near $E_\mathrm{F}$ observed in ARPES, we fitted the momentum distribution curves (MDCs) of the \CrxNbS\ data collected with LH polarization using multiple Lorentzian peaks along the cuts shown in Figure \ref{fig:MDCs}a--c. We refer to the dispersive features around $\Gamma$ near $E_{\mathrm{F}}$ as the $\alpha$, $\beta$, and $\gamma$ bands, respectively, and two parabolic bands below $E_{\mathrm{F}}$ as $\delta_1$ and $\delta_2$. Comparison of the full width at half maximum (FWHM) values from fits to the $\Gamma$--$\mathrm{M_{SL}}$ MDCs within 200 meV of $E_{\mathrm{F}}$ indicates that the two middle bands have similar FWHMs, while the outermost band (corresponding to the shallow electron pocket) is considerably broader (Figure \ref{fig:MDCs}d). We therefore assign the middle two features as split $\beta_1$ and $\beta_2$ bands in the vicinity of $E_{\mathrm{F}}$. At higher binding energies, the MDCs can be fitted well with two copies of the electron pocket bands split by about 250 meV, which we refer to as $\gamma_1$ and $\gamma_2$. Comparing the peak center positions from MDC fitting (Figure \ref{fig:MDCs}e) with the DFT band structure (Figure \ref{fig:MDCs}f) shows good qualitative agreement, other than the apparent doubling of the $\beta$ and $\gamma$ bands observed in the ARPES data.

\subsection*{Temperature Evolution of Band Structure}

\begin{figure*}[htbp]
\centering
\includegraphics[width=\textwidth]{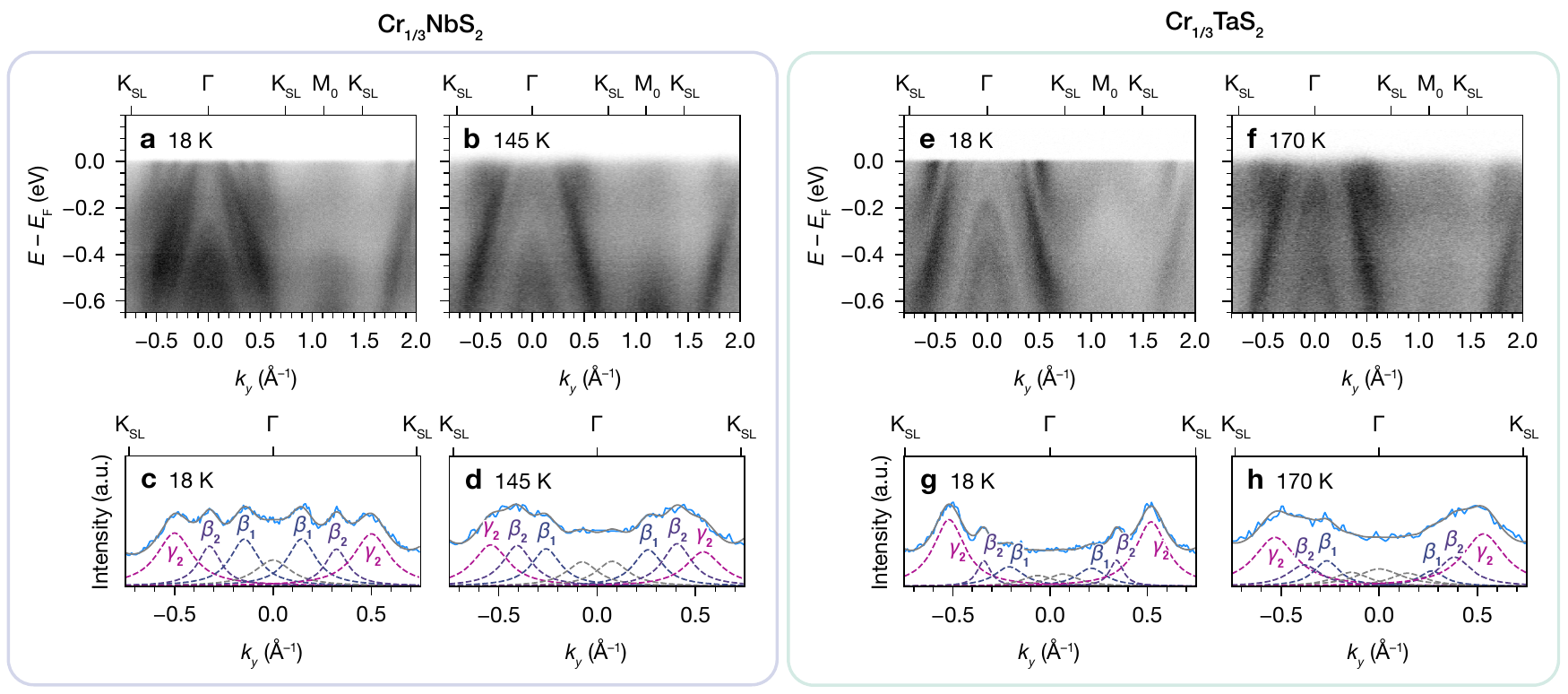}
\caption{\label{fig:temperature} (a) and (b) ARPES dispersions for \CrxNbS\ taken at 18 and 145~K, respectively. (c) and (d) Momentum distribution curves for \CrxNbS\ for $E - E_{\mathrm{F}} = -15$~meV, and fits to multiple Lorentzian peaks (dotted lines), taken at 18 and 145~K, respectively. (e) and (f) ARPES dispersions for \CrxTaS\ taken at 18 and 170~K, respectively. (g) and (h) Momentum distribution curves for \CrxTaS\ for $E - E_{\mathrm{F}} = -15$~meV, and fits to multiple Lorentzian peaks (dotted lines), taken at 18 and 170~K, respectively. All data were measured with $h\nu = 79$~eV and LH polarization.}
\end{figure*}

Due to the aforementioned band splitting, we sought to probe the effect of magnetic ordering on the electronic structures of \CrxNbS\ and \CrxTaS\ by comparing ARPES data collected below and above $T_{\mathrm{C}}$. The dispersion of \CrxNbS\ at 18~K (Figure \ref{fig:temperature}a) vs.\ 145~K (Figure \ref{fig:temperature}b) shows that the hole pockets around $\Gamma$ appear smaller at 18~K compared to 145~K. Nevertheless, multi-Lorentzian fits to the MDCs at $E - E_{\mathrm{F}} = -15$~meV show that the outer dispersive bands around $\Gamma$ crossing $E_{\mathrm{F}}$ display the same splitting at 18 and 145~K, as indicated by the labeled $\beta_1$, $\beta_2$, and $\gamma_2$ peaks in Figure \ref{fig:temperature}c--d. A similar change in hole pocket sizes is evident in the dispersions of \CrxTaS\ at 18~K (Figure \ref{fig:temperature}e) and 170~K (Figure \ref{fig:temperature}f). As with \CrxNbS, fitting the MDCs at $E - E_{\mathrm{F}} = -15$~meV indicates that the splitting of the outer bands is observed at both 18 and 170~K (Figure \ref{fig:temperature}g--h). For all the MDCs, we modeled the inner features around $\Gamma$ with Lorentzian peaks as well, but we note that ascertaining the effects of temperature on these bands is more challenging due to the lower intensities and a non-negligible background component from inelastic scattering. Nonetheless, the persistence of the $\beta_1$, $\beta_2$, and $\gamma_2$ splitting above $T_{\mathrm{C}}$ and the consistency in its magnitude for both materials prompted us to consider non-magnetic origins. 

\subsection*{ARPES Measurements with Micron-scale Probes}

\begin{figure*}[htbp]
\centering
\includegraphics[width=5.4in]{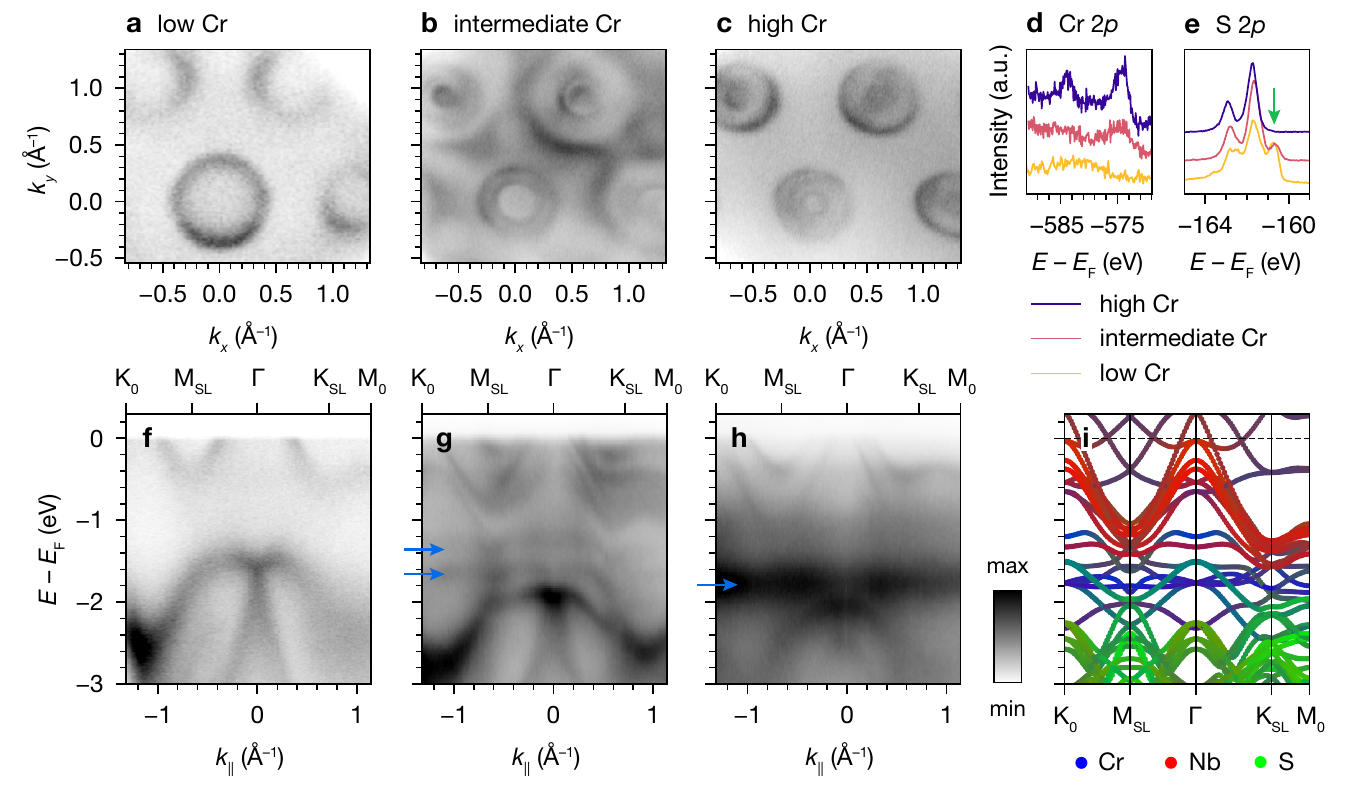}
\caption{\label{fig:surface} (a--c) ARPES Fermi surfaces of \CrxNbS\ taken in regions of low, intermediate, and high surface Cr coverage, respectively (20~K, $h\nu$ = 120~eV). (d) and (e) Cr $2p$ and S $2p$ core level spectra, respectively, with the green arrow in (e) indicating the S $2p$ feature at low binding energy. (f--h) ARPES band dispersions for the same regions with low, intermediate, and high surface Cr coverage shown in (a--c). Blue arrows in (g) and (h) indicate non-dispersive features. (i) DFT orbital-projected band structure for \CrxNbS\ showing atomic origin of bands.}
\end{figure*}

Motivated by the polar nature of these materials and the observation of unexplained band splitting, we carried out ARPES experiments on \CrxNbS\ with a smaller beam size (2--15 \textmu m) to investigate the possible impact of nonuniform sample surfaces. We identified three types of distinct areas based on their Fermi surfaces (Figure \ref{fig:surface}a--c), core level spectra (Figure \ref{fig:surface}d--e), and band dispersions (Figure \ref{fig:surface}f--h). Spots with the simplest Fermi surfaces and the largest hole pockets around $\Gamma$ and $\mathrm{K_0}$ (Figure \ref{fig:surface}a) have the weakest Cr $2p$ core level spectra (Figure \ref{fig:surface}d). Spots with Fermi surfaces representative of the majority of the samples, with the aforementioned $\beta$ and $\gamma$ band doubling (Figure \ref{fig:surface}b), exhibit Cr $2p$ core level signals of intermediate intensity. Finally, spots with Fermi surfaces missing the broadest outermost pockets around $\Gamma$ and $\mathrm{K_0}$ (Figure \ref{fig:surface}c) show the strongest Cr $2p$ core level spectra. Based on the Cr core level intensities, these areas appear to correspond to low, intermediate, and high relative Cr surface concentrations, respectively. The trend in the S $2p$ core levels from the same spots corroborate this assignment: with decreasing Cr surface coverage, an S peak at lower binding energies grows in (indicated by the green arrow in Figure \ref{fig:surface}e), consistent with more reduced S sites on the surface that are not sharing electron density with Cr.

The ARPES dispersions from these spots also exhibit notable differences. The ``low Cr'' spot (Figure \ref{fig:surface}f) exhibits less $\sqrt{3} \times \sqrt{3}$ superlattice reconstruction than the other spots (as seen from the apparent three-fold symmetry around $\mathrm{K_0}$ and different sized hole pockets at $\Gamma$ and $\mathrm{K_0}$) and resembles $2H$-\ch{NbS2} with $E_{\mathrm{F}}$ shifted up by approximately 250 meV.\supercite{elyoubi2021} The intense ``X''-shaped feature at $\Gamma$ located at about $-1.6$~eV in ``low Cr'' is shifted down to about $-2.0$~eV in both ``intermediate Cr'' (Figure \ref{fig:surface}g) and ``high Cr'' (Figure \ref{fig:surface}h) consistent with the latter two sampling more electron-doped states on average. In the ``intermediate Cr'' spot, the $\gamma$ band electron pockets near $E_{\mathrm{F}}$ are split by about 250 meV, as they are in other spectra measured with larger beam sizes. In the ``high Cr'' spot, the electron pockets near $E_{\mathrm{F}}$ are not noticeably split; instead, only a single set of features resembling the lower $\gamma_1$ band in ``intermediate Cr'' is observed. Additionally, the presence of flat bands in the ``intermediate Cr'' and ``high Cr'' spots (where they are especially prominent), as indicated by the blue arrows in Figure \ref{fig:surface}g--h, coincides with Cr states in the orbital-projected DFT band structures (Figure \ref{fig:surface}i), lending further support to the surface coverage assignments.

\section*{Discussion}

\subsection*{Band Structure and Magnetic Exchange Interactions}

Taking the results from both ARPES and DFT into account, the most pronounced difference in the band structures of \CrxNbS\ and \CrxTaS\ is the more dispersive bands in the Ta analogue. The origin appears to be steeper dispersions in $2H$-\ch{TaS2} compared to $2H$-\ch{NbS2}, i.e.\ the relative band dispersions of the host lattice materials are retained after Cr intercalation. This trend can be attributed to better overlap facilitated by more extended Ta $5d$ orbitals compared to the Nb $4d$ orbitals. For a more detailed explanation, we briefly discuss the salient bonding interactions in both host lattice materials.

\begin{figure*}[htbp]
\centering
\includegraphics[width=3.33in]{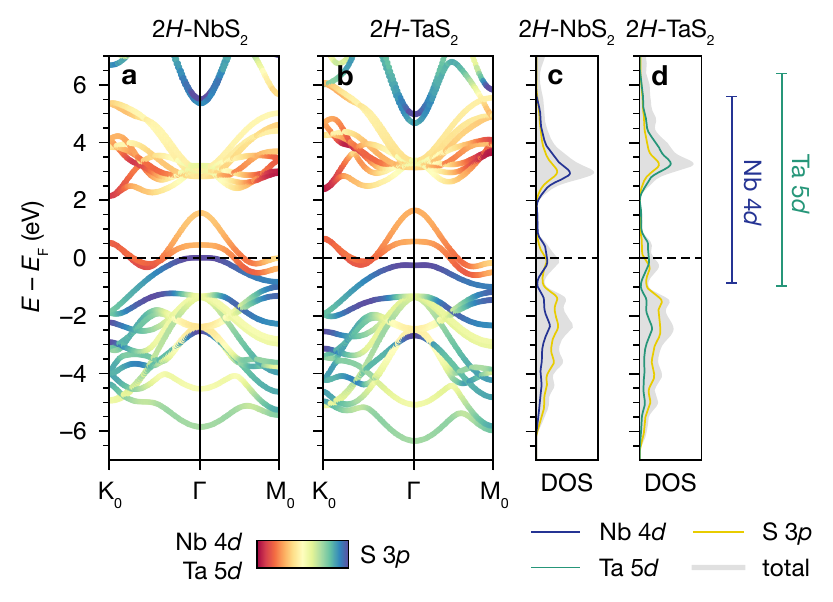}
\caption{\label{fig:hostlattice} (a) Orbital-projected band structure calculation for $2H$-\ch{NbS2}, showing Nb $4d$ and S $3p$ character. (b) Orbital-projected band structure calculation for $2H$-\ch{TaS2}, showing Ta $5d$ and S $3p$ character. (c) Projected density of states (DOS) calculation for $2H$-\ch{NbS2}. (d) Projected DOS calculation for $2H$-\ch{TaS2}. The energy spread of the Nb $4d$ and Ta $5d$ bands is indicated on the right.}
\end{figure*}

The bands within about 6 eV of the Fermi level in $2H$-\ch{$M$S2} ($M$ = Nb or Ta) are comprised of $M$ $d$ states and S $3p$ states, indicating that $M$--S $d$--$p$ and $M$--$M$ $d$--$d$ interactions are those relevant to determining the strength of the bonding and the resulting dispersivity of the bands. Mixing among the $M$ $d$ orbitals results in the formation of a hybridization gap within the $d$ manifold: the bands crossing $E_\mathrm{F}$ are composed of $d_{z^2}$, $d_{xy}$, and $d_{x^2-y^2}$ orbitals, while the higher-lying $d$ bands have more $d_{xz}$ and $d_{yz}$ character.\supercite{mattheiss1973} \ch{TaS2} has a slightly smaller in-plane lattice constant,\supercite{fisher1980,meetsma1990} and the $5d$ orbitals are more spatially extended than the $4d$ orbitals in \ch{NbS2}. This leads to better relative overlap in the Ta analogue, both in terms of $M$--S $d$--$p$ bonds and next-nearest-neighbor $M$--$M$ $d$--$d$ interactions. Hence, overall, the $d$ manifold of \ch{TaS2} is more dispersive than that of \ch{NbS2}, as shown in the band structure and density of states (DOS) calculations in Figure \ref{fig:hostlattice}b and c. In turn, the bands crossing $E_\mathrm{F}$ are also more dispersive in the Ta analogue.

These arguments can also be used to explain why the host lattice-derived bands are more dispersive in \CrxTaS\ than \CrxNbS, which we have observed in ARPES and DFT. From the crystal structures, the Ta--S bonds in \CrxTaS\ are slightly shorter (2.488(3) \AA\ on average) than the Nb--S bonds in \CrxNbS\ (2.4931(11) \AA\ on average), and the in-plane lattice constant in \CrxTaS\ is slightly smaller (5.7155(5) \AA\ vs.\ 5.7400(7) \AA), suggestive of stronger Ta--S overlap compared to Nb--S overlap. In addition, the in-plane electrical conductivity of \CrxTaS\ is more than an order of magnitude higher than that of \CrxNbS, which is consistent with the more dispersive bands and larger hole pockets in the Ta analogue expected from this analysis.\supercite{ghimire2013,obeysekera2021}

The more dispersive bands in \CrxTaS\ may have implications for the relative strengths of magnetic exchange interactions in these materials. As previously established in the literature, \CrxTaS\ has a higher $T_\mathrm{C}$ than \CrxNbS\ (133~K vs.\ 110~K for our samples). Given the extensive mixing of intercalant- and host lattice-derived states near $E_{\mathrm{F}}$ shown in DFT and supported by ARPES, our findings suggest that the magnetic exchange interactions involve some degree of itinerancy for Cr-based spins. Nevertheless, due to the considerable distance ($> 5.7$~\AA) between Cr sites, magnetic exchange is likely still mediated through TMD-derived states. The modestly higher $T_\mathrm{C}$ of \CrxTaS\ may reflect stronger FM coupling (i.e.\ larger $J_{\parallel}$ and $J_{\perp}$) in this material, which could arise from itinerant carriers with higher $v_{\mathrm{F}}$. Nevertheless, the shorter soliton wavelength and larger $H_\mathrm{c}$ imply that the ratio $D/J_{\perp}$ is considerably larger than in the Nb analogue, i.e.\ $D$ (the DM interaction term) increases more than $J_{\perp}$ in going from Nb to Ta. Higher $D$ in the Ta analogue can be ascribed to larger spin--orbit coupling (SOC),\supercite{zhang2021,obeysekera2021,du2021} and appears to affect the length and energy scales of the chiral spin textures more than changes in $J$.

The fact that \CrxNbS\ and \CrxTaS\ have qualitatively analogous magnetic phase diagrams despite different Fermi wavevectors, $k_\mathrm{F}$, supports the notion that the Ruderman--Kittel--Kasuya--Yosida (RKKY) interaction does not adequately describe exchange coupling in these materials.\supercite{ghimire2013,sirica2021} According to the RKKY formalism, the sign and magnitude of $J$ would depend closely on the magnitude of $k_\mathrm{F}$, which does not appear to be the case for \CrxNbS\ and \CrxTaS. However, the exchange interactions in other magnetic intercalated TMDs, such as \ch{Fe_{$x$}NbS2} and \ch{Fe_{$x$}TaS2}, have been proposed to be more RKKY-like.\supercite{ko2011,zheng2021,wu2022} Based on our results, we posit that \ch{TaS2}-based materials may have more dispersive bands than \ch{NbS2}-based materials for intercalants other than Cr. This difference (along with SOC) may be pertinent to the disparate magnetic properties of Nb and Ta analogues in other families of intercalated TMDs (e.g.\ \ch{Fe_{1/3}NbS2} is an antiferromagnet, whereas \ch{Fe_{1/3}TaS2} is a ferromagnet).

The most obvious route toward band engineering in \CrxNbS\ and \CrxTaS\ is through modulation of the Cr concentration. However, a delicate balance exists between disorder and vacancies in the Cr superlattice and the integrity of the desired spin textures.\supercite{dyadkin2015,kousaka2022,goodge2023} Co-intercalation of another species into the interstitial space\supercite{pan2023} or substitutional doping on the TMD sublattice could constitute other pathways toward tuning the filling level while maintaining a well-ordered Cr superlattice (and hence a globally defined $D$). Future studies in these directions could more definitively probe the effects of band dispersion on the resultant spin textures. We note additionally that the sensitivity of the observed surface states on Cr concentration---as discussed in more detail in the next section---suggests that the surface electronic structure is amenable to tuning through further functionalization.

\subsection*{Exchange Splitting vs.\ Surface Termination Effects}

Previous ARPES studies on \CrxNbS\ have also reported band splitting near $E_{\mathrm{F}}$ that appears similar to our assignment of $\beta_1$, $\beta_2$, and $\gamma_2$ bands. These works interpreted this phenomenon as exchange splitting.\supercite{sirica2016,sirica2021,qin2022} However, we observed good agreement between the exchange splitting predicted by our FM spin-polarized DFT band structures and observed in our ARPES results. Specifically, according to the spin-polarized DFT band structure calculations shown in Figure \ref{fig:dispersion}c--d, the $\alpha$ and $\beta$ bands (as labeled in Figure \ref{fig:MDCs}f) are an exchange-split pair. Because of different mixing in the spin-majority and spin-minority channels, $\alpha$ has more $d_{xy}$/$d_{x^2-y^2}$ character, while $\beta$ has more $d_{z^2}$ character (Figure \ref{fig:polarization}c and g). The corresponding bands observed in ARPES show the expected polarization dependence: the $\alpha$ band, which just touches $E_\mathrm{F}$, is much more visible in LV polarization (Figure \ref{fig:polarization}a and e), whereas the $\beta$ band is more prominent in LH polarization (Figure \ref{fig:polarization}b and f). This prompted us to consider alternative explanations for the observation of more bands in ARPES than predicted by DFT.

\begin{figure*}[htbp]
\centering
\includegraphics[width=3.33in]{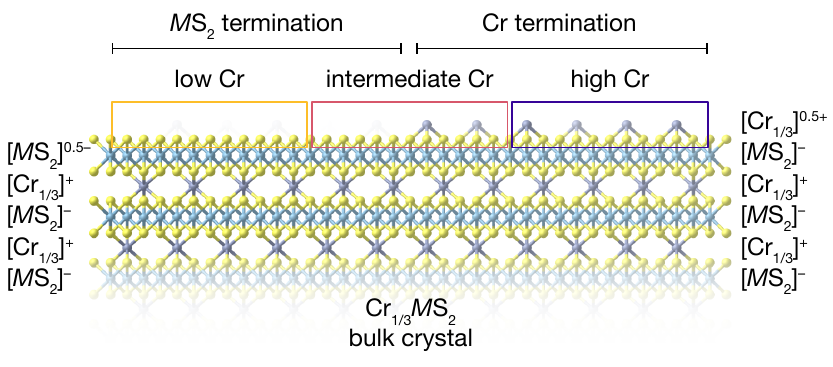}
\caption{\label{fig:cleave} Schematic illustration of local Cr clustering on cleaved surfaces of \CrxMS\ samples, with two distinct regions corresponding to predominantly \ch{$M$S2} and Cr terminations. Formal charges given for the two regions are based on a simple electron-counting picture, assuming completely full (absent) Cr coverage in the Cr- (\ch{$M$S2}-) terminated regions. The ``low Cr,'' ``intermediate Cr,'' and ``high Cr'' sampling areas (colored boxes) thus have average surface stoichiometries of 0, 1/6, and 1/3.}
\end{figure*}

Instead, taking the polar nature of intercalated TMDs into account, we surmised that the observed doubling of $\beta_1$/$\beta_2$ and $\gamma_1$/$\gamma_2$ bands could be attributed to surface termination effects. Previous work suggests that spatially distinct areas of Cr- and \ch{$M$S2}-termination exist on cleaved surfaces: STM studies have observed islands of intercalants on cleaved crystals of intercalated TMDs,\supercite{sirica2016,lim_tunable_2022} and recent ARPES studies of \ch{V_{1/3}NbS2}\supercite{edwards2023} and \ch{Co_{1/3}NbS2}\supercite{zhang2023} reported termination-dependent surface states. Thus, to understand the expected effects of surface termination on the ARPES data, we consider the charge distributions for Cr-terminated and \ch{$M$S2}-terminated surfaces. As summarized in Figure \ref{fig:overview}a, the \ch{Cr^{3+}} centers formally donate one electron per \ch{$M$S2} formula unit, leading to alternating layers with $+1$ and $-1$ formal charges. The phenomenon of charge redistribution at polar-to-nonpolar interfaces to prevent a polar catastrophe (i.e.\ diverging electrostatic potential) is well-documented,\supercite{nakagawa2006} and we expect analogous redistribution to occur at both the [\ch{$M$S2}]--vacuum interface and the [\ch{Cr_{1/3}}]--vacuum interface. Assuming fully occupied (unoccupied) Cr sites for the Cr- (\ch{$M$S2}-) terminated regions, the surface formal charges expected from simple electron counting are shown in Figure \ref{fig:cleave}. In short, \ch{$M$S2}-terminated regions should exhibit \ch{$M$S2}-derived surface states that are hole-doped relative to the bulk, originating from partial electron transfer from the surface TMD layer to compensate for its polarity.

Building upon this picture, we propose that the three distinct spectral signatures observed with mesoscopic probes (Figure \ref{fig:surface}) can be explained by sampling two different surface terminations, as illustrated in Figure \ref{fig:cleave}. The ``low Cr'' and ``high Cr'' spots correspond to areas with almost exclusively \ch{$M$S2} termination and Cr termination, respectively. The ``intermediate Cr'' spots contain both terminations, resulting in doubled $\beta$ and $\gamma$ bands that are also observed in data collected with larger probes. The $\beta$ and $\gamma$ bands at lower binding energies are associated with \ch{$M$S2}-terminated regions. Such effects are broadly consistent with those observed on surfaces of other polar layered materials.\supercite{hossain2008} Hence, we propose that these additional band doublings do not reflect magnetic ordering, but rather originate from charge redistribution. More generally, the possible contribution of surface states should be considered in other cases where unexpected bands are observed in ARPES studies on magnetic intercalated TMDs.

% ARPES primarily probes in-plane electronic structure, but the out-of-plane direction is more important for the chiral spin textures. Cr-derived bands are broader than Nb- or Ta-derived bands in both materials. Could this be a sign of electron--magnon coupling? Woolsey and White PRB 1970. Although we can't visualize the CHM state directly in ARPES, maybe there's something to be said about the low energy magnon corresponding to soliton propagation (i.e.\ trivial CHM rotation) in these materials?

\section*{Conclusions}

The electronic structures of the chiral helimagnets \CrxNbS\ and \CrxTaS\ have been investigated using ARPES and DFT. Compared to the host lattice materials $2H$-\ch{NbS2} and $2H$-\ch{TaS2}, the Cr-intercalated materials exhibit band folding from the $\sqrt{3} \times \sqrt{3}$ Cr superlattice, higher $E_{\mathrm{F}}$ from electron transfer from Cr to the host TMD, exchange splitting from the in-plane FM ordering of Cr moments, and new bands from hybridization between Cr and TMD states. The chief difference between the band structures of $2H$-\ch{NbS2} and $2H$-\ch{TaS2}---higher band dispersions in the Ta analogue---are retained after Cr intercalation, resulting in a larger $v_{\mathrm{F}}$ in \CrxTaS. This finding may have implications for the higher $T_{\mathrm{C}}$ in \CrxTaS\ via stronger FM coupling mediated by itinerant carriers.

By studying the polarization dependence in ARPES and fitting the MDCs, we find that the experimentally observed band structures agree well with the orbital-projected DFT band structures. The primary features at $E_{\mathrm{F}}$ in both materials consist of dispersive hole pockets at $\Gamma$ (and $\mathrm{K}_0$) and shallow electron pockets centered around $\mathrm{K_{SL}}$. Notably, many bands near $E_{\mathrm{F}}$ have significant Cr character in both materials, indicating that a rigid band model is insufficient for modeling the effects of Cr intercalation. Additional copies of bands crossing $E_{\mathrm{F}}$ that are not predicted by DFT are assigned to surface states originating from TMD-terminated regions. The observation of three distinct regions in ARPES experiments with smaller spot sizes is consistent with Cr, TMD, and mixed surface terminations. These results indicate that the polar nature of the surfaces of intercalated TMDs affects the band splitting observed in ARPES data.

It has been well-established that \CrxNbS\ and \CrxTaS\ have analogous magnetic phase diagrams with different energy scales and different wavelengths of magnetic solitons. Our results show that the electronic structures of these two isostructural materials are broadly analogous, with more dispersive bands in the Ta analogue. This finding suggests that band structure engineering and Fermi level tuning may allow for further modulation of the magnitude of $J$, and hence the length and energy scales of magnetic solitons in these materials. To maintain the $\sqrt{3} \times \sqrt{3}$ Cr superlattice, substitutional doping on the TMD sites or co-intercalation of other charged species may be viable routes toward controlling the properties of these chiral helimagnets, which are promising platforms for studying the interplay between electronic structure and the microscopic mechanisms underlying noncollinear magnetism.

\section*{Methods}

\subsection*{Crystal Growth}

Single crystals of \CrxNbS\ and \CrxTaS\ were grown using chemical vapor transport using iodine as a transport agent. For \CrxNbS, high-purity powders of elemental Cr, Nb, and S in a 0.6:1:2 ratio and 5 mg/cm$^3$ of \ch{I_2} were sealed under vacuum in a fused quartz ampoule approximately 48 cm long. The ampoule was placed in a three-zone tube furnace with the hot end zone and middle zone maintained at 1050 
\textdegree C and the cold (growth) zone maintained at 850 \textdegree C for 14 days before cooling to room temperature. For \CrxTaS, high-purity powders of elemental Cr, Ta, and S in a 0.47:1:2.1 ratio and 2 mg/cm$^3$ of \ch{I_2} were sealed under vacuum in a fused quartz ampoule approximately 25 cm long. The ampoule was placed in a two-zone tube furnace with the hot zone maintained at 1100 \textdegree C and the cold (growth) zone maintained at 1000 \textdegree C for 14 days before cooling to room temperature. Shiny plate-shaped crystals with a silvery metallic luster and hexagonal habit up to $4 \times 4 \times 0.5$ mm in size were obtained.

% add citations for literature precedents?

\subsection*{Structural and Compositional Characterization}

Single crystal X-ray diffraction was collected on a Rigaku XtaLAB P200 with Mo K$\alpha$ radiation at 295 K. Data reduction and scaling and empirical absorption correction were performed in CrysAlis Pro. Structures were solved by direct methods using SHELXT\supercite{sheldrick2015} and refined against $F^2$ on all data by full-matrix least squares with SHELXL.\supercite{sheldrick2015a} Raman spectroscopy was collected on a Horiba LabRAM HR Evolution with an ultra-low frequency filter using 633 nm laser excitation and powers between 1 and 8 mW. Energy dispersive X-ray spectroscopy was acquired on a FEI Quanta 3D FEG or a Scios 2 DualBeam scanning electron microscope with an accelerating voltage of 20 kV.

\subsection*{Magnetometry}

DC magnetization measurements were carried out on a Quantum Design Physical Property Measurement System Dynacool equipped with a 12 T magnet using either the Vibrating Sample Magnetometer option or the AC Measurement System II option. Single crystals were affixed to quartz sample holders with GE Varnish such that the magnetic field was applied perpendicular to the crystallographic $c$ axis.

\subsection*{ARPES Measurements}

ARPES data were collected at the Quantum Materials Spectroscopy Centre (QMSC) of the Canadian Light Source (CLS) and at Beamline 7.0.2 of the Advanced Light Source (ALS) on both the microARPES and nanoARPES endstations using Scienta Omicron R4000 hemispherical electron analyzers. The beam spot sizes were approximately $20 \times 100$ \textmu m at QMSC, $15 \times 15$ \textmu m on microARPES, and $2 \times 2$ \textmu m on nanoARPES. Results were reproduced on multiple samples at both beamlines with the exception of spatial variation observed with smaller spot sizes. Samples were cooled down to the base temperature of 20 K or below and cleaved in situ by carefully knocking off alumina posts affixed to the top surface of the sample with silver epoxy. All measurements were conducted at pressures lower than $5 \times 10^{-11}$ Torr. The primary datasets were collected at photon energies of 46, 79, and 120 eV with linear horizontal and linear vertical polarizations. Data analysis was performed using the PyARPES software package.\supercite{stansbury2020}

\subsection*{DFT Calculations}

First-principles calculations were performed by using the open source plane-wave code Quantum Espresso (QE).\supercite{QE} The optimized norm-conserving Vanderbilt (ONCV) pseudopotentials from the PseudoDojo project~\supercite{ONCV1,van2018pseudodojo} were applied. The kinetic energy cut-off for wavefunctions were set to 86 Ry for all the self-consistent calculations; for these calculations, the experimental lattice constants obtained from X-ray diffraction were used.\supercite{fisher1980,meetsma1990} A $\Gamma$-centered $4\times4\times2$ $\it{k}$-mesh was sampled in the Brillouin zone for both \CrxNbS\ and \CrxTaS, and a $8\times8\times2$ $\it{k}$-mesh for both $2H$-\ch{NbS2} and $2H$-\ch{TaS2}. The Perdew--Burke--Ernzerhof (PBE) functional~\supercite{PBE1997} of the spin-polarized generalized gradient approximation (GGA) was used to describe the exchange correlation of electrons.

\section*{Acknowledgments}

We thank Sin\'{e}ad Griffin and Ryan Day for helpful discussions, and thank Nicholas Settineri for assistance in obtaining the single crystal XRD data. The experimental work is based upon work supported by the Air Force Office of Scientific Research under AFOSR Award No.\ FA9550-20-1-0007.  L.S.X.\ acknowledges support from the Arnold and Mabel Beckman Foundation (Award No.\ 51532) and L'Or\'{e}al USA (Award No.\ 52025) for postdoctoral fellowships. O.G. acknowledges support from an NSF Graduate Research Fellowship grant DGE 1752814, and National GEM Consortium Fellowship. S.H.R.\ was supported by the QSA, supported by the U.S.\ Department of Energy, Office of Science, National Quantum Information Science Research Centers. S.H.\ acknowledges support from the Blavatnik Innovation Fellowship. Y.P.\ acknowledges the financial support from the Air Force Office of Scientific Research under AFOSR Award No.\ FA9550-21-1-0087. Confocal Raman spectroscopy was supported by a Defense University Research Instrumentation Program grant through the Office of Naval Research under Award No.\ N00014-20-1-2599 (D.K.B.). Part of the research described in this paper was performed at the Canadian Light Source, a national research facility of the University of Saskatchewan, which is supported by the Canada Foundation for Innovation (CFI), the Natural Sciences and Engineering Research Council (NSERC), the National Research Council (NRC), the Canadian Institutes of Health Research (CIHR), the Government of Saskatchewan, and the University of Saskatchewan. This research used resources of the Advanced Light Source, which is a DOE Office of Science User Facility under contract no.\ DE-AC02-05CH11231. Other instrumentation used in this work was supported by grants from the Canadian Institute for Advanced Research (CIFAR–Azrieli Global Scholar, Award No.\ GS21-011), the Gordon and Betty Moore Foundation EPiQS Initiative (Award no. 10637), the W.M.\ Keck Foundation (Award No.\ 993922), and the 3M Foundation through the 3M Non-Tenured Faculty Award (No.\ 67507585). The computational part used resources of the Center for Functional  Nanomaterials, which is a U.S.\ DOE Office of Science Facility, and the Scientific Data and Computing Center, a component of the Computational Science Initiative, at Brookhaven National Laboratory under Contract No.\ DE-SC0012704, and the lux supercomputer at the University of California, Santa Cruz, funded by NSF MRI, Grant No.\ AST 1828315, and used Stampede supercomputer at the University of Texas at Austin's Texas Advanced Computing Center (TACC) through allocation DMR160106 from the Advanced Cyberinfrastructure Coordination Ecosystem: Services \& Support (ACCESS) program, which is supported by National Science Foundation grants \#2138259, \#2138286, \#2138307, \#2137603, and \#2138296. This research was undertaken thanks in part to funding from the Max Planck--UBC--UTokyo Centre for Quantum Materials and the Canada First Research Excellence Fund, Quantum Materials and Future Technologies Program. This project is also funded by the Mitacs Accelerate Program; the QuantEmX Program of the Institute for Complex Adaptive Matter (ICAM); the Moore EPiQS Program (A.D.); and the CIFAR Quantum Materials Program (A.D.).

% \section*{Author Contributions}
% not allowed by JACS
% L.S.X., O.G., and D.K.B.\ conceived the project. O.G.\ and L.S.X.\ grew the crystals and carried out structural, compositional, and magnetic characterization with the help of C.M., S.H., and M.P.E\@. L.S.X., O.G., M.M., S.G., S.H.R., S.S.F., M.Z., N.H.J., and E.R.\ performed the ARPES experiments with assistance from S.Z., C.J., A.B., S.H., M.P.E., and A.D\@. L.S.X.\ analyzed the ARPES data with assistance from O.G., M.M., S.H.R., N.H.J., and E.R\@. K.L. carried out DFT calculations. D.K.B. and Y.P.\ supervised the work. L.S.X., O.G., K.L., and D.K.B.\ wrote the manuscript with input from all authors.

\section*{Competing Interests}

The authors declare no competing interests.

\printbibliography

%%%%%%%%%% SI %%%%%%%%%%
\pagebreak

\newrefsection  % clear references

\begin{center}
\title{\Large\textbf{Supporting Information:\\
Comparative Electronic Structures of the Chiral Helimagnets \CrxNbS\ and \CrxTaS}}

Lilia S.\ Xie, Oscar Gonzalez, Kejun Li, Matteo Michiardi, Sergey Gorovikov, Sae Hee Ryu, Shannon S.\ Fender, Marta Zonno, Na Hyun Jo, Sergey Zhdanovich, Chris Jozwiak, Aaron Bostwick, Samra Husremovi\'{c}, Matthew P.\ Erodici, Cameron Mollazadeh, Andrea Damascelli, Eli Rotenberg, Yuan Ping, D.\ Kwabena Bediako

\end{center}

% Prefix a "S" to all equations, figures, tables and reset the counter
\setcounter{equation}{0}
\setcounter{figure}{0}
\setcounter{table}{0}
\setcounter{page}{1}
\makeatletter
\renewcommand{\thepage}{S\arabic{page}}
\renewcommand{\theequation}{S\arabic{equation}}
\renewcommand{\thefigure}{S\arabic{figure}}
\renewcommand{\thetable}{S\arabic{table}}

\section*{Single Crystal X-ray Diffraction}

\begin{figure}[htbp]
\centering
\includegraphics{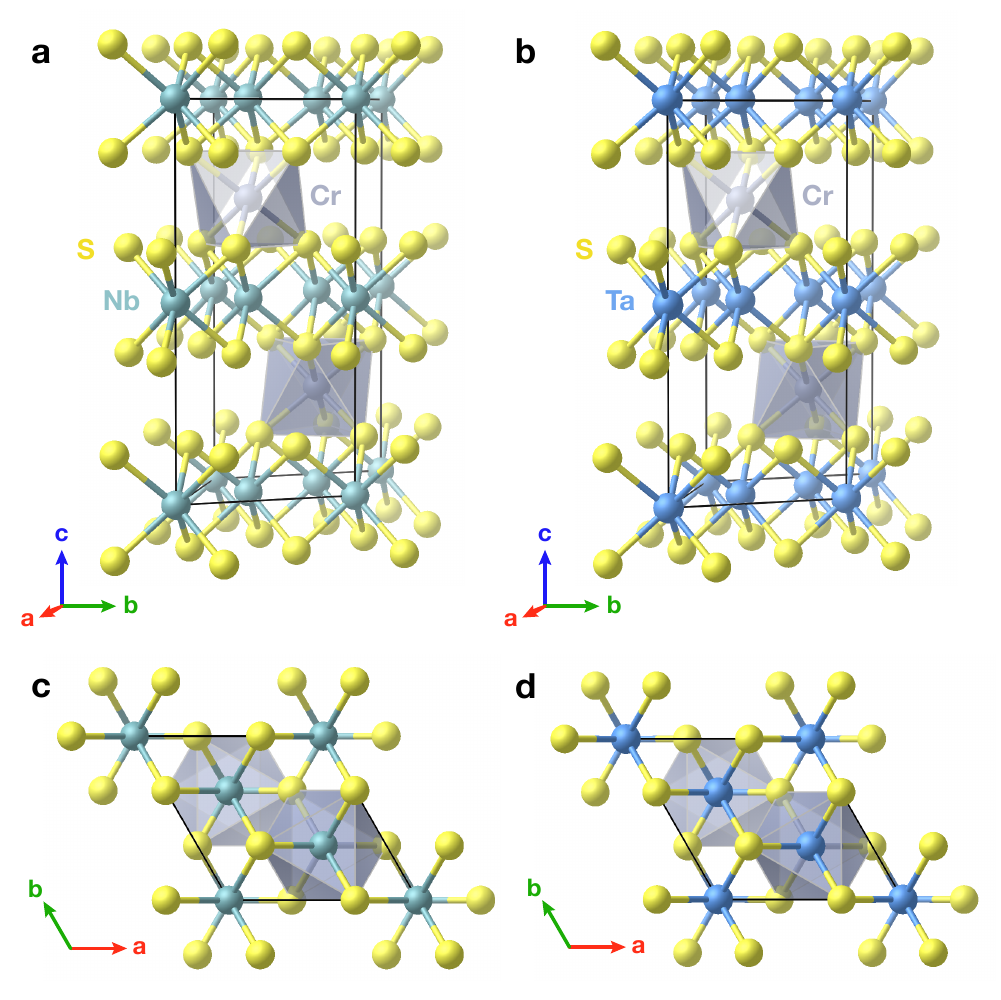}
\caption{\label{fig:structure}Crystal structures of \CrxNbS\ and \CrxTaS\ from single crystal X-ray diffraction. Representations are shown for \CrxNbS\ and \CrxTaS, respectively, along the $a$ crystallographic axis in (a) and (b), and $c$ crystallographic axis in (c) and (d).}
\end{figure}

\begin{table}[htbp]
\footnotesize
    \caption{Crystal data and structure refinement for \CrxNbS\ and \CrxTaS.}
    \centering
    \begin{tabular}{lll}\toprule
         & \CrxNbS & \CrxTaS\\\midrule
         Empirical formula & \ch{CrNb3S6} & \ch{CrTa3S6}\\
         Formula weight (g/mol) & 523.09 & 787.21\\
         Temperature (K) & 293(2) & 293(2)\\
         Wavelength (\AA) & 0.71073 & 0.71073\\
         Crystal system & Hexagonal & Hexagonal\\
         Space group & $P6_322$ & $P6_322$\\
         $a$ (\AA) & 5.7400(7) & 5.7155(5)\\
         $c$ (\AA) & 12.1082(14) & 12.1751(12)\\
         Volume (\AA$^{-3}$) & 345.49(9) & 344.44(7)\\
         $Z$ & 2 & 2\\
         Density (calculated) (g/cm$^3$) & 3.402 & 5.680\\
         Absorption coefficient (mm$^{-1}$) & 8.082 & 50.733\\
         $F(000)$ & 486 & 678\\
         Crystal size (mm$^3$) & $0.033 \times 0.017 \times 0.013$ & $0.119 \times 0.067 \times 0.025$\\
         $\theta$ ($\degree$) & 3.365 to 29.272 & 4.117 to 29.531\\
         Index ranges & $-6 \leq h \leq 7$ & $-7 \leq h \leq 7$\\
         & $-7 \leq k \leq 6$ & $-7 \leq k \leq 7$\\
         & $-14 \leq l \leq 16$ & $-16 \leq l \leq 16$\\
         Reflections collected & 2684 & 2641\\
         Independent reflections & 297 & 302\\
         Completeness to $\theta_\mathrm{full}$ & 1.000 & 0.994 \\
         Absorption correction & Semi-empirical from equivalents & Semi-empirical from equivalents\\
         Refinement method & Full-matrix least-squares on $F^2$ & Full-matrix least-squares on $F^2$\\
         Data / restraints / parameters & 297 / 0 / 17 & 302 / 0 / 17\\
         Goodness-of-fit on $F^2$ & 1.312 & 1.156\\
         Final $R$ indices [$I > 2\sigma(I)$] & $R_1$ = 0.0331, $wR_2$ = 0.0848 & $R_1$ = 0.0381, $wR_2$ = 0.1166\\
         $R$ indices (all data) & $R_1$ = 0.0435, $wR_2$ = 0.0880 & $R_1$ = 0.0437, $wR_2$ = 0.1213\\
         Largest diff. peak and hole ($e$\,\AA$^{-3}$) & 1.81 and $-0.65$ & 4.83 and $-1.89$\\\bottomrule
         
    \end{tabular}
    \label{tab:crystal}
\end{table}

\begin{table}[htbp]
\footnotesize
    \caption{Atomic coordinates, Wyckoff positions, and equivalent isotropic displacement parameters for \CrxNbS.}
    \centering
    \begin{tabular}{llllll}\toprule
        Atom Labels & $x$ & $y$ & $z$ & Site & $U_{\mathrm{iso}}$\\\midrule
        Cr01 & 2/3 & 1/3 & 3/4 & $2c$ & 0.0067(6)\\
        Nb02 & 0 & 0 & 1/2 & $2a$ & 0.0037(4)\\
        Nb03 & 1/3 & 2/3 & 0.50283(6) & $4f$ & 0.00236(3)\\
        S04 & 0.6680(3) & 0.6675(3) & 0.63086(11) & $12i$ & 0.0048(4)\\\bottomrule

    \end{tabular}
    \label{tab:CrxNbS2}
\end{table}

\begin{table}[htbp]
\footnotesize
    \caption{Atomic coordinates, Wyckoff positions, and equivalent isotropic displacement parameters for \CrxTaS.}
    \centering
    \begin{tabular}{llllll}\toprule
        Atom Labels & $x$ & $y$ & $z$ & Site & $U_{\mathrm{iso}}$\\\midrule
        Cr01 & 2/3 & 1/3 & 3/4 & $2c$ & 0.0098(7)\\
        Ta02 & 0 & 0 & 1/2 & $2a$ & 0.0047(5)\\
        Ta03 & 1/3 & 2/3 & 0.50232(4) & $4f$ & 0.0048(5)\\
        S04 & 0.6679(3) & 0.6685(3) & 0.6304(2) & $12i$ & 0.0058(7)\\\bottomrule

    \end{tabular}
    \label{tab:CrxTaS2}
\end{table}

\pagebreak

\section*{Raman Spectroscopy}

\begin{figure*}[htbp]
\centering
\includegraphics{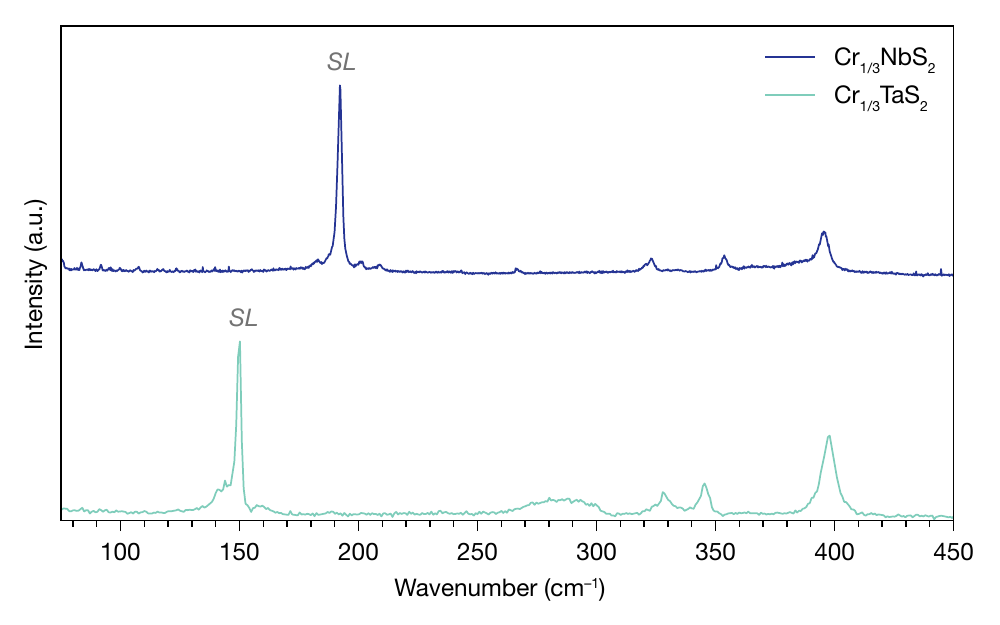}
\caption{\label{fig:raman} Raman spectra of \CrxNbS\ and \CrxTaS, with modes associated with the $\sqrt{3} \times \sqrt{3}$ superlattice labeled as ``SL'' and modes associated with the host lattice materials labeled according to symmetry.\supercite{fan2021}}
\end{figure*}

\pagebreak

\section*{Energy Dispersive X-ray Spectroscopy}

\begin{figure*}[htbp]
\centering
\includegraphics{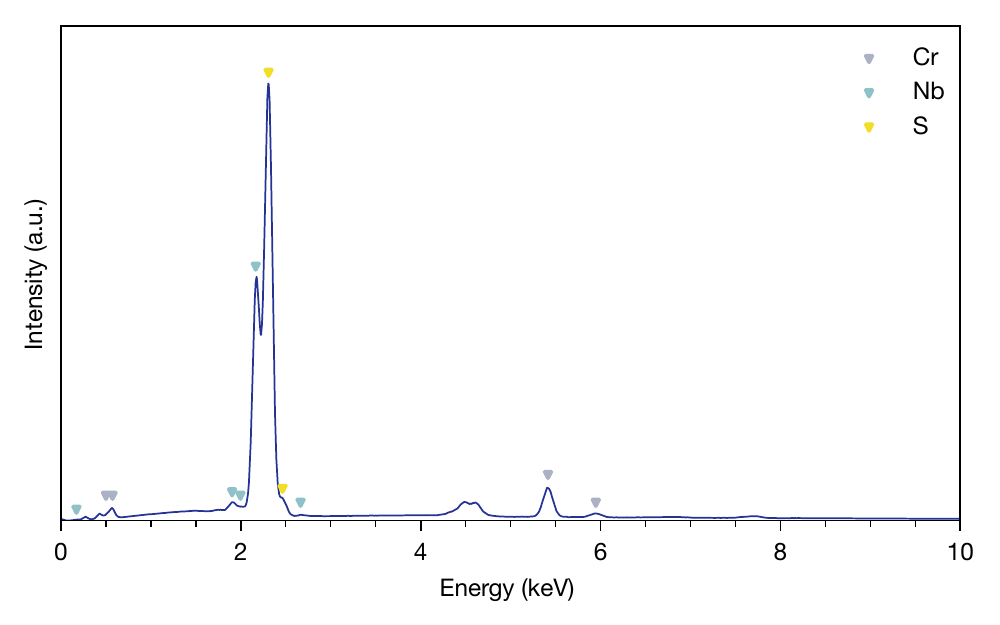}
\caption{\label{fig:EDS_Nb} Representative dispersive X-ray spectroscopy data for a single crystal of \CrxNbS\ with peaks corresponding to Cr, Nb, and S labeled. The atomic ratio determined by fitting the Cr K$\alpha_1$, Nb L$\alpha_1$, and S K$\alpha_1$ peaks was 1.00:3.00:6.30., corresponding to a formula of \ch{Cr_{0.33}NbS_{2.10}}.}
\end{figure*}

\begin{figure*}[htbp]
\centering
\includegraphics{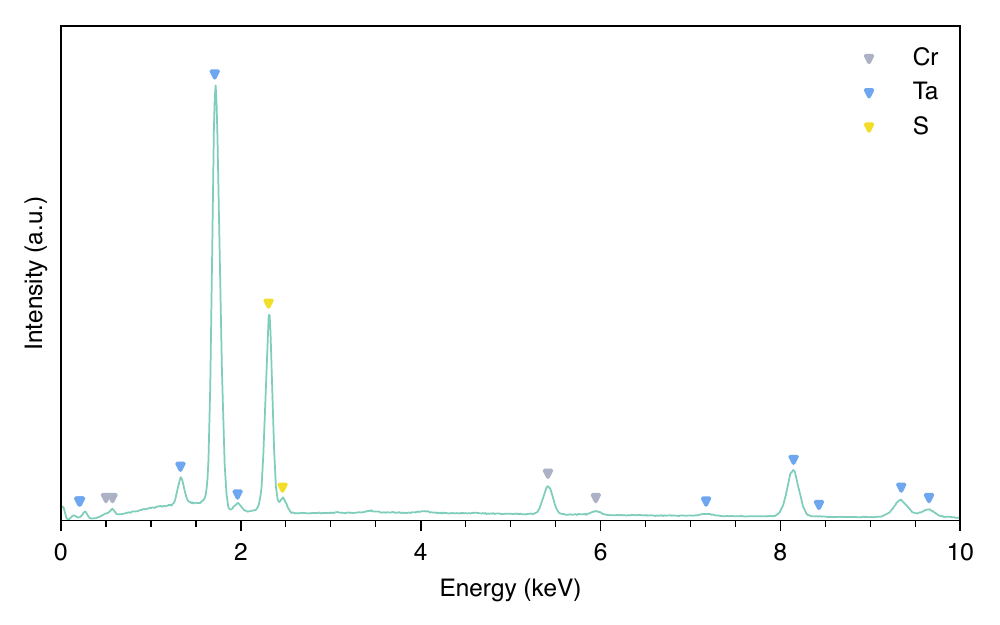}
\caption{\label{fig:EDS_Ta} Representative dispersive X-ray spectroscopy data for a single crystal of \CrxTaS\ with peaks corresponding to Cr, Ta, and S labeled. The atomic ratio determined by fitting the Cr K$\alpha_1$, Ta M$\alpha_1$, and S K$\alpha_1$ peaks was 1.00:3.07:5.70, corresponding to a formula of \ch{Cr_{0.33}TaS_{1.86}}.}
\end{figure*}

\pagebreak

\section*{Magnetometry}

\begin{figure*}[htbp]
\centering
\includegraphics{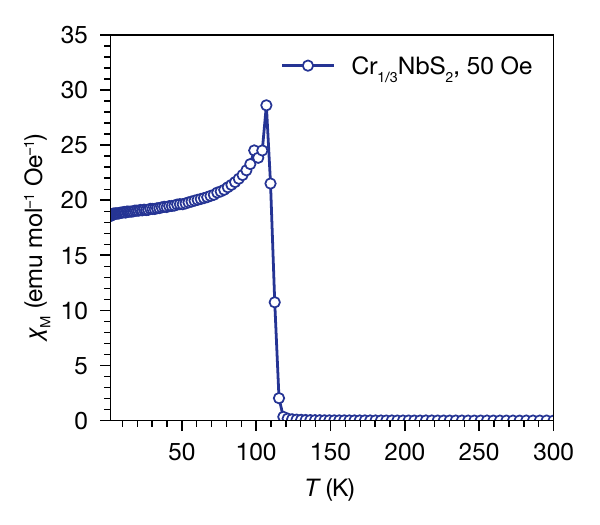}
\caption{\label{fig:MT_Nb} Temperature-dependent magnetic susceptibility for \CrxNbS, measured with $H \perp c$ = 50 Oe.}
\end{figure*}

\begin{figure*}[htbp]
\centering
\includegraphics{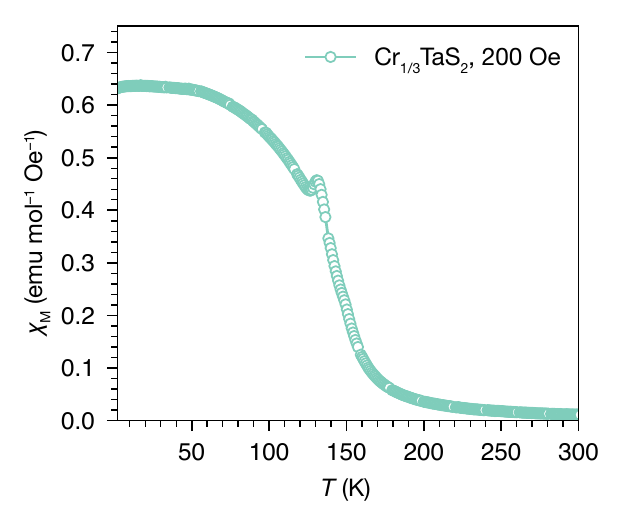}
\caption{\label{fig:MT_Ta} Temperature-dependent magnetic susceptibility for \CrxTaS, measured with $H \perp c$ = 200 Oe.}
\end{figure*}

\pagebreak

\section*{DFT Calculations}

To understand the band structures obtained from ARPES measurements, first-principles calculations for pristine $2H$-\ch{NbS2}, $2H$-\ch{TaS2}, and the Cr-intercalated analogs of these host lattices were performed by using the open source plane-wave code Quantum Espresso (QE).\supercite{QE} The optimized norm-conserving Vanderbilt (ONCV) pseudopotentials from the PseudoDojo project~\supercite{ONCV1,van2018pseudodojo} were applied. The kinetic energy cut-off for wavefunctions were set to 86 Ry for all the self-consistent calculations; for these calculations, the experimental lattice constants obtained from X-ray diffraction were used. A $\Gamma$-centered $4\times4\times2$ $\it{k}$-mesh was sampled in the Brillouin zone for both \CrxNbS\ and \CrxTaS, and a $8\times8\times2$ $\it{k}$-mesh for both $2H$-\ch{NbS2} and $2H$-\ch{TaS2}. The Perdew--Burke--Ernzerhof (PBE) functional~\supercite{PBE1997} of the spin-polarized generalized gradient approximation (GGA) was used to describe the exchange-correlation of electrons. Previous studies incorporating first-principles calculations at the PBE level\supercite{ghimire2013,bornstein2015,sirica2021} and at the GGA+$U$ level~\supercite{qin2022} used an on-site Coulomb interaction, $U$, of 4 eV for Cr. In this work, different $U$ parameters for Cr, Nb, and Ta were explored. The results obtained from PBE are shown and compared to the experiments in the main text. Calculations obtained for the spin-polarized, orbital-projected, $k_z$-dependent band structures, and the Fermi surfaces are shown in comparison with the results.

\subsection*{\textbf{\textit{U}} Parameter for Band Structure from GGA+\textbf{\textit{U}} Calculations}
Cr, Nb, and Ta are transition metals for which the $U$ parameter can be used to describe the on-site Coulomb interaction between localized $d$ electrons.\supercite{shi2009magnetism,csacsiouglu2011effective} However, whether or not the $U$ parameters for the aforementioned transition metals are important for the band structure calculations of \CrxNbS\ and \CrxTaS\ is not clear. It needs more investigation beyond the previous studies using PBE\supercite{ghimire2013,bornstein2015,sirica2021} and the study using GGA+$U$ with a $U$ value of 4 eV for Cr.\supercite{qin2022} Here, the on-site Coulomb interaction $U$ parameters for Cr and Nb are tested using the values close to those from Ref.~\cite{shi2009magnetism,csacsiouglu2011effective,pasquier2018charge,calandra2018phonon,qin2022}. The $U$ parameters are discussed in order to clarify the effect of $U$ parameters on the band structure and the necessity of adopting $U$ parameters. 

\begin{figure}[htbp]
    \centering
    \includegraphics[width=\textwidth]{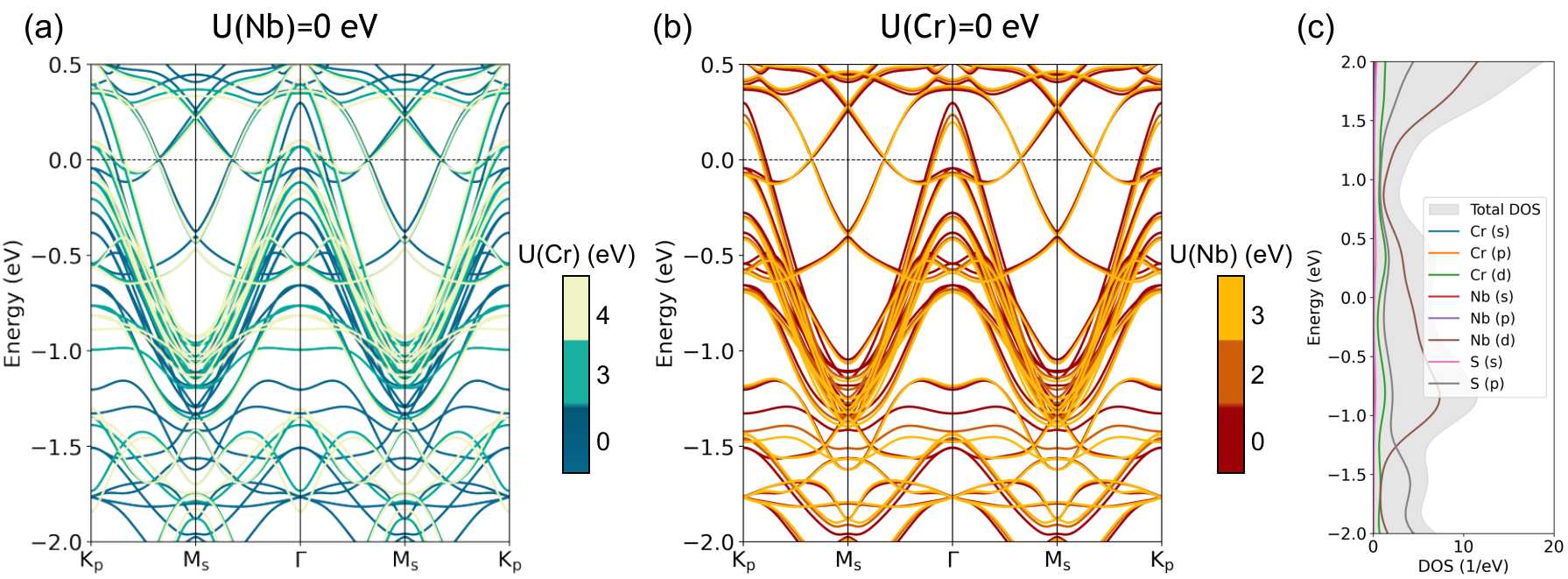}
    \caption{(a) and (b) Band structures of \CrxNbS\ with the on-site Coulomb interaction for Cr and Nb represented as $U$(Cr) and $U$(Nb), respectively. (c) density of states of \CrxNbS\ calculated with PBE. (a) $U$(Nb) is set to 0 eV and the effect of $U$(Cr) on the band structure is shown when varied from 0 to 4 eV. (b) $U$(Cr) is set to 0 eV and the effect of $U$(Nb) on the band structure is shown when varied from 0 to 3 eV.}
    \label{fig:bandstruct_crnbs2_vs_U}
\end{figure}

Looking at the band structures with varying $U$(Cr) in Figure~\ref{fig:bandstruct_crnbs2_vs_U}a, overall the band structures of interest within $-$1 eV to 0 eV do not change much with $U$(Cr). Moreover, the band structures near the Fermi level do not show qualitative changes, except for the minor upshift of the bands at $\Gamma$ with increasing $U$(Cr). %One effect that could be noticed is the band structure within -1 eV to 0 eV upshifting about 0.05 eV by every 1 eV increase of U(Cr). 
The small change of this part of band structures may be explained by the fact that $d$ orbitals of Cr are minor in the composition as shown in Figure~\ref{fig:bandstruct_crnbs2_vs_U}c of the partial density of states (PDOS), and that the $d$ electrons near the Fermi level are nearly delocalized and thus not affected by the $U$ parameter. Likewise, $U$(Nb) can be found to be not important to the band structures of interest from Figure~\ref{fig:bandstruct_crnbs2_vs_U}b. Thus, the band structures from PBE are shown in the main text and compared with experiments.

\begin{figure}[htbp]
    \centering
    \includegraphics[width=\textwidth]{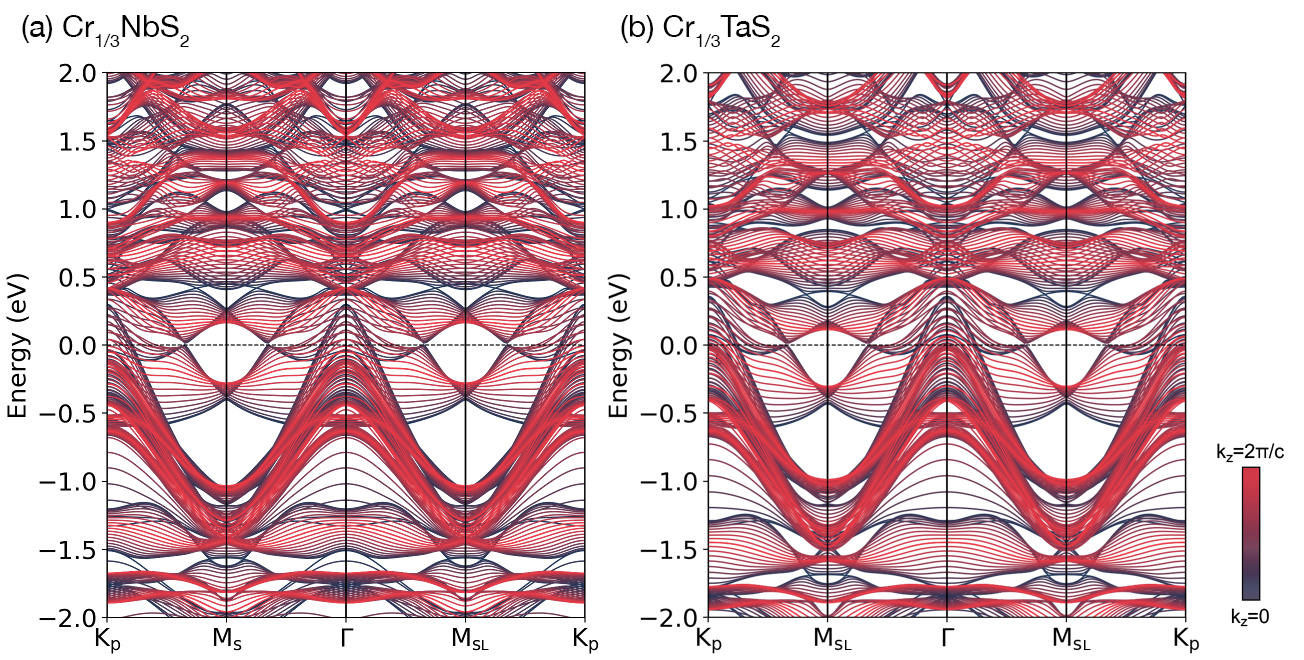}
    \caption{(a) and (b) $k_z$-projected band structures of \CrxNbS\ and \CrxTaS, respectively, calculated using the PBE functional.}
    \label{fig:slab}
\end{figure}

\clearpage

% note from LX: the value calculated from Curie-Weiss fits is mu_eff, which is not necessarily the same as the saturation magnetization. Same w/ the other types of measurements... In order to make sure we're comparing apples to apples, I think it may be better to only include saturation magnetization values measured from M(H) in these tables.

\begin{table}[ht]
    \footnotesize
    \centering
    \caption{Magnetic moment of \CrxNbS.}
    \begin{tabular}{ccccc}
        \hline
        Stoichiometry & Saturation magnetization ($\mu_\text{B}$) & Reference & Notes\\
        \hline
        1/3 & 2.9 & \cite{miyadai_magnetic_1983} & $M(H)$ (4.2 K) \\
        % 1/3 & 1.5 & \cite{togawa_chiral_2012-1} & (expt.) Value is from TEM measurements at 110 K\\
        1/3 & 3.89 & \cite{hulliger1970magnetic} & $M(H)$ (2 K) \\
        % 1/3 & 3 & \cite{chapman2014spin} & (expt.)\\ this one is a theory paper, they just assumed mu B = 3 right off the bat
        1/3 & 3.2 & \cite{ghimire2013} & $M(H)$ (2 K)\\
        % 1/3 & 4.4 & \cite{braam2015magnetic} & (expt.) Obtained from muon spin rotation \\
        % & & & and relaxation measurements\\
        % 1/3 & 2.9(1) & \cite{mayoh2022giant} & (expt.) Calculated from Curie Weiss law fits\\
        % 1/3 & 3.94(13) & \cite{hall2022} & (expt.) Calculated from Curie Weiss law fits\\
        0.33(1) & 2.68 & This work & $M(H)$ (2 K)\\
        % 0.33(1) & 3.77 & Our data & (expt.) Same sample as above using \\
        % & & & Curie-Weiss fits of the linear regime\\
        % & $3.2\pm0.9$ & & (expt.) Average\\
        \hline
        1/3 & 2.66 & This work & DFT (PBE)\\
        \hline
    \end{tabular}
    \label{tab:magn_crnbs}
\end{table}

\begin{table}[ht]
    \footnotesize
    \centering
    \caption{Magnetic moment of \CrxTaS.}
    \begin{tabular}{ccccc}
        \hline
        Stoichiometry & Saturation magnetization ($\mu_\text{B}$) & Reference & Notes\\
        \hline
        % 1/3 & 3.91 & \cite{parkin1980} & (expt.)\\ this one also curie-weiss
        1/3 & 2.97 & \cite{zhang2021} & $M(H)$ (2 K)\\
        1/3 & 2.73 & \cite{obeysekera2021} & $M(H)$ (2 K)\\
        1/3 & 2.73 & \cite{du2021} & $M(H)$ (2 K)\\
        % 1/3 & 2.9 & \cite{yamasaki2017} & (expt.) Measured using transport from flakes\\
        % 1/3 & 2.73 & \cite{obeysekera2021} & (expt.) Calculated from Curie Weiss law fits\\
        0.33(1) & 2.82 & This work & $M(H)$ (2 K)\\
        \hline
        % & $2.6\pm1.2$ & & (expt.) Average\\
        1/3 & 2.71 & This work & DFT (PBE)\\
        \hline
    \end{tabular}
    \label{tab:magn_crtas}
\end{table}

\pagebreak

\printbibliography

\end{document}